\documentclass[12pt,nofootinbib,notitlepage,aps,prd,tightenlines]{revtex4-1}
\usepackage{amsmath,amssymb}
\usepackage{verbatim,graphicx}
\usepackage{multirow}
\usepackage{hyperref}
\usepackage{array}
\usepackage{color}

\newcommand\vev[1]{\left\langle #1\right\rangle}

\DeclareMathOperator{\tr}{\text{tr}}
\DeclareMathOperator{\Imag}{\text{Im}}
\DeclareMathOperator{\Real}{\text{Re}}
\newcommand\CO{{\cal O}}

\begin{document}
\setlength{\unitlength}{1mm}


\title{Large $N$ classical dynamics of holographic matrix models}



\author{Curtis T. Asplund}
\email{curtis@itf.fys.kuleuven.be}
\affiliation{Institute for Theoretical Physics, KU Leuven, 3001 Leuven, Belgium}
\altaffiliation{Department of Physics, University of California, Santa Barbara, CA 93106}

\author{David Berenstein}
\email{dberens@physics.ucsb.edu}

\author{Eric Dzienkowski}
\email{eric@physics.ucsb.edu}
\affiliation{Department of Physics, University of California, Santa Barbara, CA 93106}

\date{\today}

\begin{abstract}
Using a numerical simulation of the classical dynamics of the plane-wave and flat space matrix models of M-theory, we study the thermalization, equilibrium thermodynamics and fluctuations of these models as we vary the temperature and the size of the matrices, $N$. We present our numerical implementation in detail and several checks of its precision and consistency. We show evidence for thermalization by matching the time-averaged distributions of the matrix eigenvalues to the distributions of the appropriate Traceless Gaussian Unitary Ensemble of random matrices. We study the autocorrelations and power spectra for various fluctuating observables and observe evidence of the expected chaotic dynamics as well as a hydrodynamic type limit at large $N$, including near-equilibrium dissipation processes. These configurations are holographically dual to black holes in the dual string theory or M-theory and we discuss how our results could be related to the corresponding supergravity black hole solutions.

\end{abstract}
\maketitle

\section{Introduction}
\label{sec:intro}

Some of the greatest successes of the holographic dualities have been the computations of transport coefficients of strongly coupled quantum field theories, beginning with the work of \cite{Policastro:2001yc}. 
These computations replace a problem of large $N$ (high dimensional gauge group), finite temperature, strongly coupled quantum gauge theory dynamics by the much simpler problem of analyzing classical Green's functions in various black hole backgrounds with boundary conditions that reflect our understanding of response theory. 

The direct calculation of the transport coefficients in the strongly coupled quantum field theory is currently beyond reach.
The main problem is that response theory at finite frequency requires us to solve the real time evolution of an interacting quantum system of many degrees of freedom.
If this is done in a path integral by Monte-Carlo sampling, the dynamical situation we want to consider has a severe sign problem that is currently unsolved.
In some instances there are workarounds (for a recent review see \cite{deForcrand:2010ys}).

The calculation in  \cite{Policastro:2001yc} and subsequent work is, to leading order and when properly normalized \cite{Kovtun:2004de}, completely independent of $N$ and all the precise details of the field theory. 
These are classical gravity computations in AdS black hole backgrounds. 
Corrections appear when the gravitational theory suffers quantum corrections, i.e., when the curvature is large somewhere in Planck units, or when the truncation to gravity breaks down, as when the black hole has curvature radii of order the string scale. 
In the dual field theory, this corresponds to finite $N$ or weak coupling. 
In such cases it may pay off to approach the problem from the other side of the duality, starting with a field theory Hamiltonian and studying its dynamics and then examining its relation to the gravity limit. This paper falls into this category of work.

Given our inability to solve strongly coupled quantum theories directly, we could instead try to drop the quantum adjective and solve ``strongly coupled classical theories.'' 
At first, this might seem impossible, but we have to consider that strong coupling in quantum theories also refers to the strength of the terms in the action that make the system non-linear, relative to the terms corresponding to a linear system \footnote{In perturbation theory in quantum mechanics this can also be the regime where energy denominators are large, but off-diagonal terms in the perturbation are even larger, so the standard conditions for perturbation theory to make sense do not necessarily apply.}.
Thus, the regime of interest in classical physics is that of a dynamical system in the very non-linear regime. 
We also want to study it at large $N$. Here we expect that the strong non-linearity will cause chaotic dynamics and that large $N$ should be understood as a thermodynamic limit, where the number of degrees of freedom grows as $N^2$ but some aspects of the dynamics are $N$-independent. 
Beyond the existence of good thermodynamic state variables, we can look for collective modes to emerge, akin to hydrodynamic variables, that indicate collective time dependent dynamics also roughly independent of $N$. 

Classical non-linear field theories have an infinite number of degrees of freedom and suffer from the ultraviolet (UV) catastrophe.
The UV catastrophe is cured by reintroducing the Planck constant.
Indeed, this is how Planck introduced the eponymous constant in the first place, giving birth to quantum mechanics.
This would seem to stop this idea of studying classical non-linear field theory dynamics in its tracks. 
However, thinking of a field theory expanded in Fourier modes, a finite $\hbar$ freezes the dynamics of most of the modes to be in their ground state (or their adiabatic ground state given a configuration of the low frequency modes). 
These frozen modes are those above a cutoff, which is determined by the dynamics and the initial conditions.
So, one is only dealing with finitely many active degrees of freedom and the initial problem of an infinite number of degrees of freedom in field theory can be solved, if we only knew the precise details of how most of the degrees of freedom ``freeze out." 

This problem is solved without any work if we study dynamical systems with finitely many degrees of freedom in the first place.
In that case, we don't need to address the UV catastrophe at all.
In those systems, $\hbar \to 0$ means that we are studying the system at large quantum numbers (very large energies compared to the gap in systems with a discrete spectrum).
This is a regime where typical states are well described by classical statistical mechanics.

With this in mind, our main purpose is to examine the real time classical statistical mechanics of certain matrix models that have finitely many degrees of freedom. 
This is exactly the same methodology as in molecular dynamics simulations (see, e.g., \cite{MolecDyn} for a review).
We study both equilibrium configurations, with their associated equations of state, and also simple transport processes, or more precisely, out of equilibrium relaxation. 
We study the latter via fluctuations of the appropriate variables and by invoking the fluctuation-dissipation theorem.
We discuss whether such classical models can be used to study holographic dualities, where we also have some gravitational information. 
That is, to what extent can the classical dynamics of large $N$ matrix models encode gravitational information? 
More broadly, are the dynamics compatible with our expectations from string theory, including beyond the general relativity/supergravity regime?
We are not able to completely answer these questions, but do make some relevant qualitative comparisons and, especially in Sec. \ref{sec:grav-int}, offer some informed speculations on the full story, which is a subject of continued research.

We work with the BFSS \cite{BFSS} and BMN \cite{BMN} matrix models, which have well known holographic dual descriptions. 
We simulate the real time classical dynamics of these models. 
In our simulations the generic late time behavior for the initial conditions we study are finite temperature, equilibrium configurations.
For the reasons we discussed above, it is reasonable to treat these holographic matrix models classically at high temperatures, which is the regime of very large quantum numbers.
The associated black holes have curvatures that are large in string units \cite{Itzhaki:1998dd}.
Thus, from the gravitational perspective we would be describing very stringy back holes with our analysis. 

We also find that there are collective dynamical variables analogous to hydrodynamic degrees of freedom. 
That is, there are certain aspects of the near equilibrium dynamics that are independent of the number of microscopic degrees of freedom.
This was not obviously guaranteed. 
Hydrodynamics usually requires a geometric coarse graining of degrees of freedom confined to small regions.
Holographic systems are usually made of D-branes. 
The holographic degrees of freedom are the strings stretching between the branes, which are extended objects that generally do not localize to small volumes. 
In the case of AdS bulk geometries, even small regions on the AdS boundary have infinitely large volumes in AdS, so this issue does not apply directly.
One does hydrodynamics on the boundary of AdS and not in the bulk, as in \cite{Policastro:2001yc}.  
In the case of the BMN matrix model, the conformal boundary of the dual plane wave geometry has no spatial extent \cite{Berenstein:2002sa, Marolf:2002ye} and so one cannot have transport there.
Hence, hydrodynamic behavior is not an obvious possibility in this case.
 
However, the membrane paradigm for black holes suggests that there should be transport phenomena for the degrees of freedom in the near horizon limit of our black hole type objects. 
We still have to fret about the possibility of a phase transition: that the classical regime of the matrix model (at large $N$) has nothing to do with the black hole dynamics we would like to study.
In \cite{Horowitz:1996nw} the authors argued that the transition from branes or strings to black holes is a smooth process, which also suggests some form of collective behavior in the near horizon dynamics, but it might be very distorted from the classical gravity regime. 
See also \cite{Festuccia:2006sa}, which also argues against any such phase transition at any finite $N$ or finite 't Hooft coupling. 

We are also motivated to study this problem by the fast scrambling conjecture of \cite{Sekino:2008he} and the hope that we can find some numerical handle to study it that does not involve gravitational arguments in the first place.
We are continuing the study that was initiated in \cite{Asplund:2011qj}, where some notion of fast classical scrambling was shown for the BMN matrix model. 
Another analysis in the BFSS matrix model was carried out in \cite{Riggins:2012qt}, which also showed fast classical thermalization in the BFSS matrix model for some set of initial conditions. 
However, these studies are not in the same spirit as the Sekino-Susskind setup, which seems to fundamentally require $\hbar \neq 0$ in its formulation. 
Unfortunately, we have nothing new to say in regards to this conjecture, even though the principal example that was argued to thermalize fast was exactly the BFSS matrix model. The crucial property of the model, shared by the ones we study, was that all of its degrees of freedom are coupled to each other in the Lagrangian. 
Indeed, it has been argued in \cite{Edalati:2012jj} that non-commutative field theory, which makes the degrees of freedom more non-local, scrambles faster than the corresponding commutative version of the field theory dynamics, but this was again done via a gravitational computation.

Other attempts to study the fast scrambling conjecture directly include \cite{Lashkari:2011yi}, which studies certain toy quantum mechanical models, but not holographic systems. 
See the series of papers \cite{Barbon:2011pn, Barbon:2011nj, Barbon:2012zv} for a more detailed treatment of holographic fast scramblers, though these do not study matrix models directly. 
The papers \cite{Festuccia:2006sa, Iizuka:2008hg, Iizuka:2008eb} slightly predate the conjecture and study matrix model dynamics and thermodynamics perturbatively at weak-coupling, as well as the breakdown of perturbation theory, 
and so are complementary to this work.

The paper is organized as follows.
In Sec. \ref{sec:obsandsym} we discuss the regime where our calculations are valid, and we discuss the classes of observables that can be considered as well as a discussion on what it means for such  matrix model to behave hydrodynamically.
In Sec. \ref{sec:num} we describe the algorithm we use to evolve configurations. Of particular importance is that the dynamics requires a Gauss' law constraint, and our checks that the constraint is preserved to high accuracy.
In Sec. \ref{sec:thermal} we show that the late time behavior of the matrix model dynamics seems to thermalize and we present tests of this property, we also solve a puzzle on how to compute the temperature, where two independent measurements at first glance seem to disagree.
In Sec. \ref{sec:chaos} we compute the power spectra of interesting observables, especially in the BFSS matrix model. We show that the autocorrelation functions of interesting observables have  smooth power spectrum (indicative of chaos) and  have a well defined large $N$ limit suggesting hydrodynamics behavior.
In Sec. \ref{sec:fac} we study the large $N$ factorization of the statistical quantities of interest.
We show numerically that in the classical dynamics various quantities behave to leading order as gaussians (free fields) and that a $1/N$ expansion is applicable to study various normalized correlation functions.
Finally we conclude.

\section{Observables and symmetry}
\label{sec:obsandsym}

The objective of this paper is to study the classical dynamics of both the BFSS and BMN matrix models and to see to what extent we can extract lessons about holography from this study. 
Before we formulate our approach to that problem, we discuss the regime where our calculations are valid.

The regime where the four-dimensional $\mathcal{N} = 4$ SYM theory is dual to a semi-classical (super) gravity theory is large $N$ and strong 't Hooft coupling \cite{Maldacena:1997re}.
The strong coupling regime implies that $g^2_{\text{YM}}N\hbar \gg 1$, so it involves $\hbar$ in a  crucial way.
One should be careful in interpreting this equation. 
In dimensions other than four, like our $0+1$ dimensional matrix models, the Yang-Mills coupling constant has units, so the left hand side can not be compared to the right hand side without choosing a state and multiplying by appropriate quantum numbers to obtain a dimensionless ratio.

If we choose a thermal state at temperature $T$ for an oscillator degree of freedom with angular frequency $\omega$, we can ask if thermal fluctuations are larger than quantum fluctuations for that degree of freedom. 
This happens when $k_\text{B} T \gg \hbar \omega$.
So, one can be in the small $\hbar$ regime if the temperature is high enough. 
For relativistic quantum field theories there is always some $\omega$ where $k_\text{B} T < \hbar \omega$. Such degrees of freedom would be responsible for the UV catastrophe. 
On the other hand, for oscillators with small $\omega$ the left hand side is much larger than the right hand side and the corresponding oscillators are at high occupation quantum numbers. 
The dynamics of these low frequency modes is controlled by classical physics. 
The classical world meets the quantum world in the intermediate regime.  
Roughly speaking, the correspondence principle in quantum mechanics should let us interpolate between the classical and the quantum regimes. 
In the BMN and the BFSS matrix model we only have a finite number of degrees of freedom, so the UV catastrophe issue is avoided, but we can still try to push the system to the correspondence limit, in a manner we describe below.

Typical quantum states are superpositions of position eigenstates, so if we are to match various physical quantities of the quantum system we should either average over positions or smear the classical states to a volume of $\hbar$ for each canonical pair of variables. 
If the system is chaotic, most energy eigenstates behave as if they are thermal for sufficiently small subsystems \cite{Srednicki} (one has to make allowances if there are conserved quantities which don't thermalize), so the correspondence principle suggests that we should study the statistical properties of the thermal ensemble to study the coarse grained properties of the quantum states. 
In this paper we do not calculate anything in the regime where quantum effects start making a difference, but we keep in mind that in the end we want to understand the system in the quantum regime.

%

Now, let us go back to the BMN and BFSS matrix models.
The regime of interest for us is the large $N$ regime, as dictated by holography.  
In this regime, the number of degrees of freedom grows like $N^2$. 
We will see that the large $N$ limit is not only a thermodynamic limit but that we also observe a kind of hydrodynamics.
Here we don't mean thermodynamic limit in a formal, rigorous sense, but simply that we find various state variable that remain finite as $N$ grows.
Ideally, we should be able to show local equilibrium and transport to claim hydrodynamic behavior.
Unfortunately, we don't know how to make such a formulation from first principles, as the degrees of freedom in these matrix models are essentially non-local.
If we take one of the matrices of the matrix model, we can interpret the eigenvalues as positions of D-branes \cite{Dai:1989ua}, while the off-diagonal elements are strings stretching between the branes.
In the classical regime we are studying, all the off-diagonal modes are excited, so it is hard to define local quantities that could play the role of, e.g., densities of D-particles.

What we can do instead is add a faraway probe and determine what it sees. 
In the BFSS matrix model such a formulation leads to an effective potential for the probe.
The effective potential can be computed from traces of the configuration \cite{BFSS} convolved with some Green's functions that decay polynomially in the distance \cite{Taylor:1998tv}.  
This is a quantum computation where one integrates out the off-diagonal degrees of freedom connecting the probe to the configuration under study.
This gives rise to gravitational interactions between general D-brane objects and gravitons (the interactions between gravitons by integrating off-diagonal models is part of the original formulation of the BFSS matrix model \cite{BFSS} \footnote{The simplest one loop computation was done in \cite{DKPS,BC}, while a two loop result was obtained in \cite{Becker:1997xw}.
Higher order results require information on the wave functions of the graviton states one is scattering.}).
The natural candidates for hydrodynamic variables are these traces of various products of matrices appearing in the effective potential.
They act as moments of the distributions of matter in the effective potential for a faraway probe. 

We need to show that the dynamics of these collective modes are roughly independent of $N$ to justify calling their dynamics hydrodynamic.
Note that we only make claims about these collective degrees of freedom in a statistical sense, just as if we were considering the hydrodynamic variables of a system of molecules. 
Since we will be studying mostly equilibrium configurations, all we have access to is the fluctuations of these variables.
Our results show these fluctuations have some dynamical properties independent of $N$.

Specifically, we will study time dependent correlation functions of certain single trace observables. 
Consider first a single trace operator 
\begin{equation}
\CO_{[i]} = \tr \left( X^{i_1} X^{i_2} \dots \right),
\end{equation}
where $[i]$ is a multi-index.
In the brane picture this will be a source for some gravity field (or more generally a closed string field).
Indeed, we usually find that 
\begin{equation}
\CO_{[i]} \simeq \int d^dx\, \rho_\alpha(x) f(x),
\end{equation}
where $\rho_\alpha$ would be the local source of the field (if such a notion makes sense), with its corresponding spin labels.
This is convolved with some polynomial function of the $x$, which we call $f$, which can also carry angular momentum labels.
Together these would get combined into a multipole expansion labeled by the multi-index $[i]$. 
We can decompose the product into spherical harmonics, and then symmetry considerations will tell us that if the configuration is spherically symmetric the averages of objects whose spin is non-zero vanishes.
For many interesting observables we will have that the time average should vanish
\begin{equation}
\langle \CO_{[i]} \rangle_t = 0,
\end{equation}
although it does not do so configuration by configuration, but only as a time average.

Now, if we give two such observables $\CO_i(t)$, we can consider averages of the form
\begin{equation}
S_{ij}(a) = \langle \CO_i(t) \CO_j(t+a) \rangle_{t}
\end{equation}
where we average over a trajectory (or a collection of such trajectories with the same energy).
The correlation function $S_{ij}(a)$ will describe the statistical properties of the time dependent correlations between the observables.
Indeed, such a function will encode the fluctuation-dissipation information.
Such correlation functions can be different from zero, even if the individual expectation values of the $\CO_i$ vanish. 
This is similar to studying sound modes for gas in a cavity.
If the individual harmonics are not excited, then their average is zero, but there will be thermal fluctuations.
These fluctuations, when properly normalized, will 
have a good thermodynamic limit,
but away from this limit there can be finite size effects that are sensitive to the number of particles.

We will say the system behaves hydrodynamically if a collection of the $S_{ij}(a)$ properly normalized has a large $N$ limit where the collection $S_{ij}(a)$ converges to a single function of $a$ for fixed $ij$, and for a reasonable interval of time that is short compared to the Poincar\'e recursion time, but that can be much longer than any thermalization time (or scrambling or relaxation time) for near equilibrium dynamics.

This may be too narrow a definition.
Consider a toy model of gas in a box with a somewhat random shape that is temperature dependent (like a rubber balloon filled with an ideal gas).
We would say that the hydrodynamic behavior there is independent of the box.
However, let us imagine that we want to study hydrodynamics by looking at the normal modes of sound in the box, or other such decomposition into normal modes.
If we change the temperature, we would change the shape of the box somewhat: the added pressure would deform the walls of the container. 
This would deform the harmonics of the box, and the collection of harmonics of the box would be temperature dependent.
Such changes cannot be done while preserving the spectrum (even after rescaling time).
Also, if we change the number of particles inside the box in such a way that the pressure stays the same, the shape of the box would not change.
However, the temperature of the gas would change depending on the number of particles.
In such a case, the modes of sound on the box would be independent of the temperature only after a rescaling of time, and the ratios of the different frequencies would be invariant, but not the frequencies themselves.

Our systems are somewhat analogous to this.
After thermalization, the matrices will relax to an approximately spherically symmetric configuration about the center of mass.
This is in the absence of angular momentum for the initial conditions (we will not consider such initial conditions in this paper). 
These spherical configurations grow in size if we increase the temperature.
In the BMN system the geometry of the plane wave in which the configuration is embedded acts like a box, similar to how AdS acts like a mirror. 
If we increase the temperature the configuration grows in size and the external pressure changes. 
There is also an internal pressure that makes the system want to collapse: the excitation of the off-diagonal modes between the branes acts like a glue that makes the system shrink. 
If these are treated as harmonic oscillators, one would expect that each such harmonic oscillator has an energy of $k_\text{B}T$ and that the energy stored in these configurations is independent of the position. 
However, as we move a D-particle far away from the system, the effective frequency of these modes goes up, and there is a corresponding shrinkage of the available phase space for these modes.
Thus, there is an entropy cost to move a D-particle away from the configuration and the internal pressure to hold the system together is an entropic force. 
It has been argued that this type of entropic effect leads to the gravitational force near the horizon of a black hole \cite{Verlinde:2010hp}.

On the other hand, thermal pressure makes the system expand.
These two forces can reach an equilibrium.
In the very high temperature limit we expect that the internal pressure dominates over the external pressure, so that the shape of the container matters less, but we will not be able to guarantee that the system is hydrodynamic without fine tuning: we would need to be able to match the box shapes between different values of $N$.
Doing this carefully requires a fairly detailed understanding of the phase diagram of the system.

All of this is much simpler to study in the BFSS matrix model. One can show that the classical dynamics of the BFSS model 
has a scaling property: a configuration at a given energy can be rescaled to any other energy by a rescaling of time and the matrix components.
Thus, it does not matter at what energy we study the system as the dynamics is essentially the same. So we have gotten rid of the temperature-dependent shape parameters. 
Because of this, we can argue that the dynamics of the thermal system only depends on $N$ and we can explore how the dynamics depends on $N$ and no other variables.
For of this reason, we analyze the large $N$ limit primarily in the BFSS matrix model.

\section{Numerical implementation}
\label{sec:num}

The numerical implementation has been discussed previously in \cite{Asplund:2011qj}.
We work with a leapfrog algorithm and we indicate how to implement the constraints in the initial conditions.
Here we reiterate the algorithm.

The bosonic degrees of freedom of the BMN and BFSS matrix model are the hermitian matrices $X^{i=0,1,2}$ and $Y^{a=1, \ldots, 6}$ and their canonical conjugates $P_i$ and $Q_a$.
The bosonic part of the Hamiltonian is
\begin{align}
H &= \frac{1}{2}\tr \Big(P_i^2 + Q_a^2 + (\alpha X^i + i\epsilon^{ijk} X^j X^k)^2 \nonumber \\
&\quad \hskip 1cm + \frac{\alpha}{4}(Y^a)^2 - [X^i, Y^a]^2 - \frac{1}{2}[Y^a, Y^b]^2 \Big). \label{H}
\end{align} 
When $\alpha \neq 0$ we have the BMN matrix model.
For convenience we set $\alpha = 1$ in the BMN case.
For $\alpha = 0$ we get the BFSS matrix model.

We have rescaled the variables so that the classical equations of motion are independent of $\hbar$ and all the quantum mechanics is hidden in the initial conditions.
We have also normalized the mass of $X$ to one, \emph{i.e.}, we measure time by the oscillation period of one of the $X$ modes. 

Because of the $\mathrm{U}(N)$ gauge symmetry we must enforce the Gauss' law constraint:
\begin{equation}
C =  [X^i, P_i] + [Y^a, Q_a] = 0\,. \nonumber
\end{equation}
To solve the equations of motion we use a leapfrog algorithm.
This has the virtue of preserving the constraints.
The discretized matrix equations of motion read
\begin{equation}
\label{eq:eom}
X_{t + \delta t} = X_t + P_{t + \frac{\delta t}{2}} \delta t \,, \qquad
P_{t + \frac{\delta t}{2}} = P_{t - \frac{\delta t}{2}} - \frac{\partial V}{\partial X}\Big |_t \delta t\, 
\end{equation}
and similarly for the $Y$ modes.
Since we have the $X^i$, $P_i$ evaluated at different times, we need to be a little careful with the constraint.
We define 
\begin{equation}
C(t) =  [X^i(t), P_i(t+\delta t/2)] + [Y^a(t), Q_a(t+t/2)]
\end{equation}
and will set it to zero in the initial conditions.
We also define the constraint at half intervals to be given by
\begin{equation}
C(t + \delta t / 2) = [X^i(t + \delta t), P_i(t + \delta t / 2)] + [Y^a(t + \delta t), Q_a(t + \delta t/2)].  
\end{equation}
To show that the constraints are satisfied notice that when we evolve the constraint by using the equation of motion \eqref{eq:eom}, after one half step in $t$ we get that
\begin{align}
C(t + \delta t / 2) - C(t) &= \sum_i [\delta X^i(t), P_i(t + \delta t / 2)] + \dots \nonumber \\
&= \sum_i [P_i(t + \delta t / 2), P_i(t + \delta t / 2)] \delta t + \dots = 0,
\end{align}
which vanishes term by term.
For the second half step we need to work harder, but so long as $V$ is a sum of traces of products of $X$ and $Y$ matrices (or functions of such traces), one can prove that the contribution from each such trace vanishes by summing cyclically over the letters making the word in the trace.
Hence, $C(t + \delta t) - C(t) = 0$ and this tells us that $C(t)$ is a constant of motion of the discrete evolution.
Incidentally, the same arguments work for angular momentum conservation laws.
Our initial conditions are those for (near) zero angular momentum.
The only place where constraint violations might appear is from rounding errors, so we need to check that we don't suffer from this problem.
To improve numerical stability we use double precision numbers. 
To check for numerical errors, we record the absolute value of the constraint $CV = |\tr(C^2)|$ as a check for the code.
We find that the constraint is well satisfied for the runs we perform, so we do not need to implement constraint damping.
The equations of motion evolve hermitian matrices into hermitian matrices.
Truncation errors in matrix multiplication can also take us away from that locus.
We found that we needed to enforce hermiticity of the matrices every few steps, by taking $X \to (X + X^\dagger) / 2$ and similarly for all other matrices.
We do this every time we write the matrix configurations. 

The main sources of difficulty in the setup are the initial conditions.
For this paper, we have used the following initial classical configurations:
\begin{align}
\label{initialcond}
X^0 &= \begin{pmatrix}L^0_{n} & 0 \\
0 & 0 \end{pmatrix}, \,  \,
X^1 = \begin{pmatrix} L^1_{n} & \delta x_1 \\
\delta x_1^\dagger & 0 \end{pmatrix}, \,   \,
X^2 = \begin{pmatrix} L^2_{n} & \delta x_2 \\
\delta x_2^\dagger & 0 \end{pmatrix}, \nonumber \\
P_0 &= \begin{pmatrix} 0 & 0 \\
0 & v \end{pmatrix}, \quad P_{1,2} = 0 = Q_{1, \dots, 6}, \quad
Y^a = \delta y^a.
\end{align}
in the BMN matrix model.
These are the same initial conditions that were used in \cite{Asplund:2011qj}, where they are explained in detail.
The important point is that the $\delta x$ and $\delta y$ are generated by Gaussian distributions controlled by a classical estimate of the quantum uncertainty of the modes which is parametrized by $\hbar$.
This is only used in the initial conditions and is similar to the standard practice in molecular dynamics simulations 
\cite{MolecDyn}.

To study the BFSS matrix model, we first evolve in the BMN matrix model and some time later we set $\alpha = 0$ and study the further evolution of the system.
Once we are in the BFSS matrix model, we have further dynamical information: the classical system enjoys a scale invariance.
Thus, results at one energy are completely equivalent to results at higher or lower energies (there is no temperature dependence on quantities, other than those dictated by scaling) and the only variable we have is $N$.
This makes it easy to compare different values of $N$ to each other after proper rescaling.
In this situation we can test convergence of quantities as we increase $N$ in a temperature independent way.

We store the full configurations of the matrices every few steps in $\delta t$ (for the data sets we present here we set this number to ten unless otherwise stated), and we store other information for faster processing at different intervals.
This is especially important for long simulations.
We will call machine time the total run in the simulation in units of the smallest time step that is recorded. 

\begin{figure}[hb]
\includegraphics[width=2.5 in]{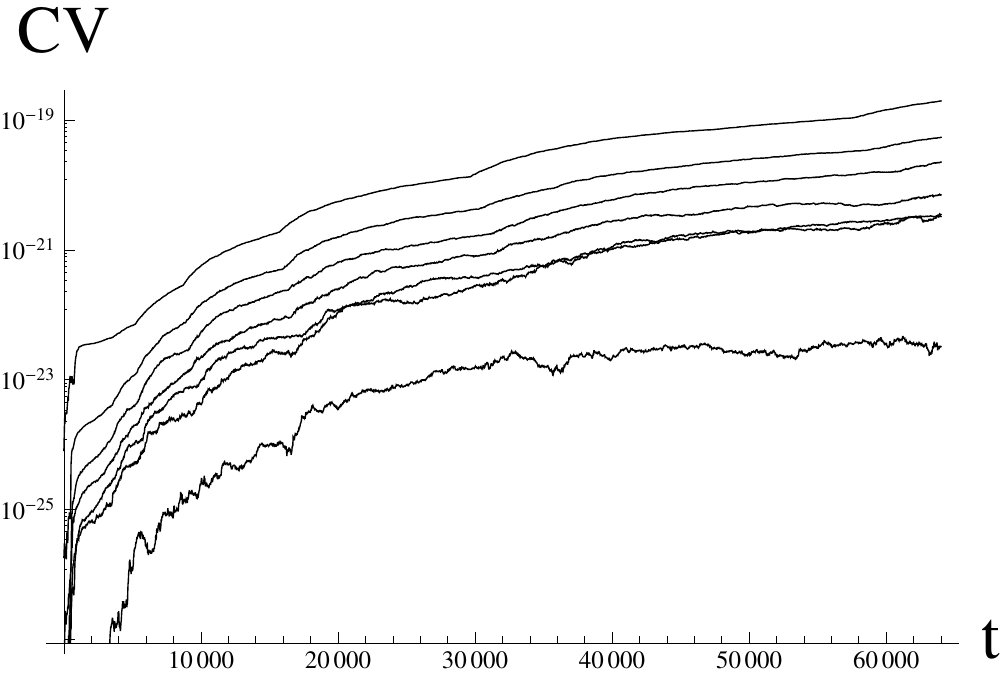}\includegraphics[width=2.5 in]{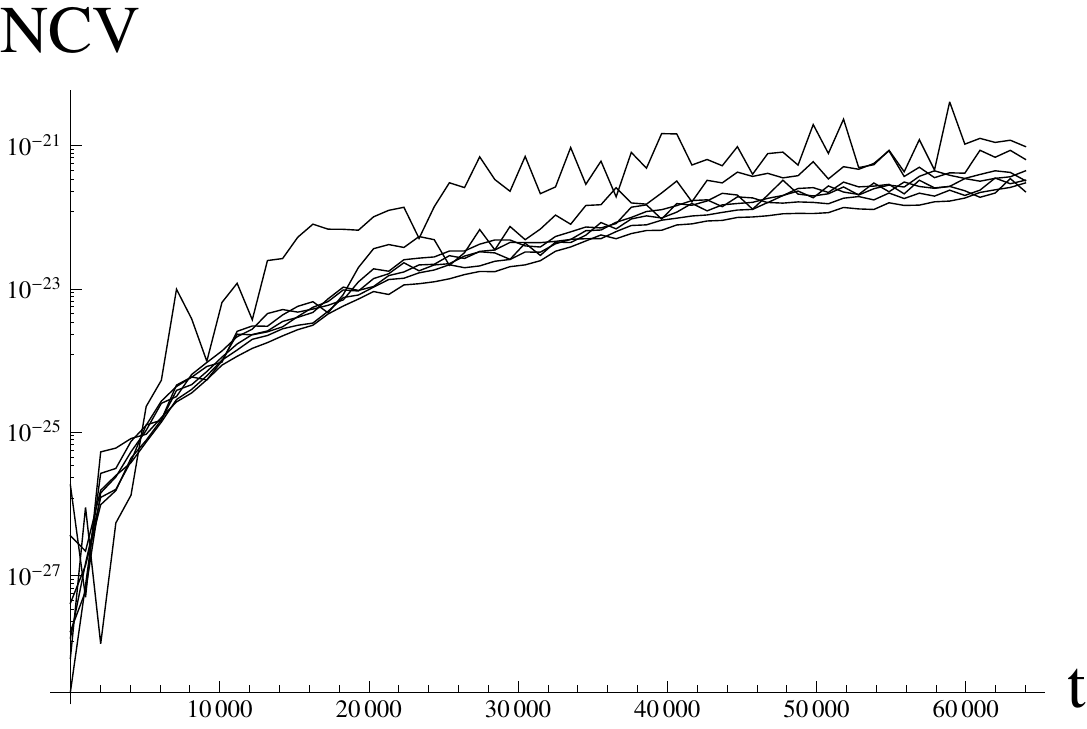}
\caption{Constraint violation as a function of machine time and normalized constraint violation.
As can be seen the constraint violation stays very low through the computation.
The graph indicates values of $N=4,7,10,13,18,27,47$, and after normalization it shows larger fluctuations for smaller $N$.
These are due to the statistical fluctuations of $\tr (X^2)\tr(P^2)$, which are an integral part of the dynamics.
In the plot to the right we have selected larger intervals in machine time to aid visualizing the different values of $N$.}
\label{fig:conv}
\end{figure}

Our results for the constraint violation can be seen in Fig. \ref{fig:conv}.
As the constraints technically have units, we need to normalize them, and we define the normalized constraint violation to be given by $NCV = -N \tr(C^2) / (\tr(X^2)\tr(P^2))(t)$ (we randomly chose $X^2$, $P_1$) and $N$ being the size of the matrices.
In the simulations depicted in the figure \ref{fig:conv} we set $\alpha = 0$ after machine time $t = 2000$.

\section{Thermalization}
\label{sec:thermal}

\begin{figure}[b]
\includegraphics{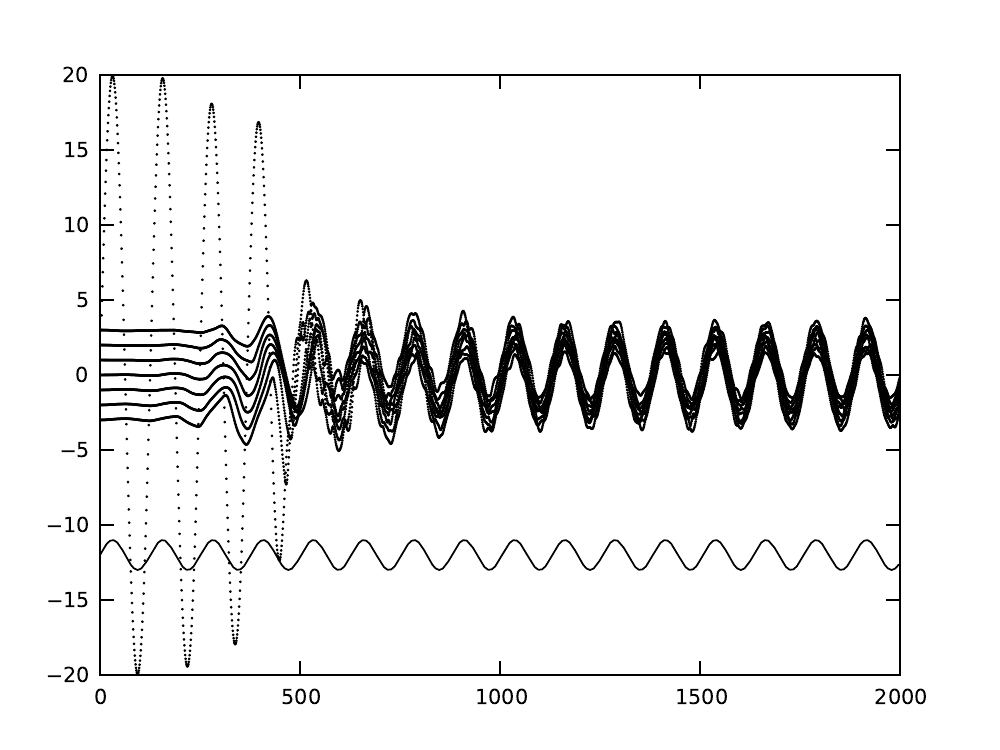}
\caption{Eigenvalues of $X^0$ as a function of time for a simulation of rank 8 matrices with $\delta t = 0.001$, $v = 20.0$, and $\hbar = 0.001$.
The abscissa measures discrete time units between recordings.
The sinusoidal curve at the bottom is the trace of $X^0$ which serves as a clock due to its equations of motion.}
\label{eigenblob}
\end{figure}

In \cite{Asplund:2011qj} it was shown that the initial conditions (\ref{initialcond}) generate configurations of eigenvalues which coalesce into a uniformly oscillating blob, for example see Fig. \ref{eigenblob}.
The system was argued to have thermalized in that time averaged distributions of the momentum degrees of freedom follow a Gibbs ensemble $dP\, dQ \, \exp(-\beta H)$ for some inverse temperature $\beta$.
The Gibbs distribution factors into a product of Gaussians because both the BFSS and the BMN Hamiltonians are quadratic in the momenta.
It was shown that the binned eigenvalues collected over time follow the semicircular distribution for random matrices, as is expected for random Gaussian matrices in the large $N$ limit.
Here we explore in more detail the nature of the thermalization, the appropriate finite $N$ Gibbs distribution, and the design of a correctly calibrated thermometer.

One thing to note is that the equations of motion for the traces of the coordinates and their momenta are those of a harmonic oscillator:
the trace of the $X^i$ and $P^i$ oscillate with period $2\pi$ while the $Y^a$ and $Q^a$ oscillate with period $4\pi$.
Although this property of the BMN system can be used to generate a clock for the system (as in Fig. \ref{eigenblob}), it is undesired here.
The trace is a protected degree of freedom that does not thermalize.
To describe the thermalized system in a statistical manner, we must remove the trace degree of freedom from our Gibbs distribution.
The matrices are Hermitian, and so we must also enforce that on our Gibbs distribution.
What we are left with is the Traceless Gaussian Unitary Ensemble (TGUE), a means by which to select traceless random Hermitian matrices.
The trace of the matrices represents the center of mass motion of the system and thus our partition function really only describes the internal degrees of freedom that can thermalize.

In order to study ensemble quantities of the system, we must coarse grain the dynamics leaving only gauge invariant quantities to be studied.
We can study our Gibbs distribution in a gauge invariant manner by focusing on eignvalues and traces.
Integrating over the unitary degrees of freedom gives the joint probability distribution for the eigenvalues.
This result is well known for the GUE and is simple to modify for the traceless case in which we are interested.
The trace is invariant under a unitary transformation, and thus we may enforce tracelessness by inserting a delta function without affecting the removal of the unitary degrees of freedom.
We define the following partition function as the integral of the joint probability of the eigenvalues for the TGUE
\begin{equation}
\label{thermalization:eigentgue}
\mathcal{Z}_{\lambda} = \int d^N\lambda\, \delta(\tr(\Sigma_i \lambda_i))\prod_{1\leq i < j \leq N}|\lambda_i - \lambda_j|^2 \exp\left( -\frac{\beta}{2}\sum_{i=1}^N \lambda_i^2\right)
\end{equation}
where $N$ is the rank of the matrices of interest and $\beta$ is a parameter analogous to the standard deviation for normal distributions.
For us $\beta$ is physically the inverse temperature of our system.
Note that the polynomial in the integrand is the square of the Vandermonde determinant of the eigenvalues.

\begin{figure}
\includegraphics[width=8cm]{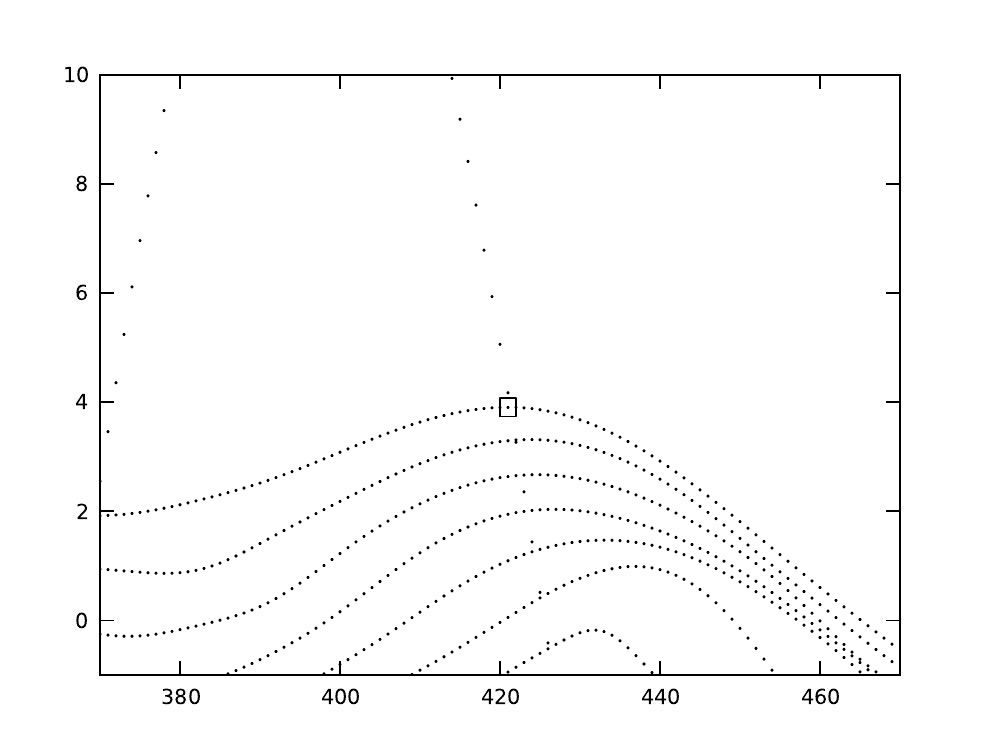}
\includegraphics[width=8cm]{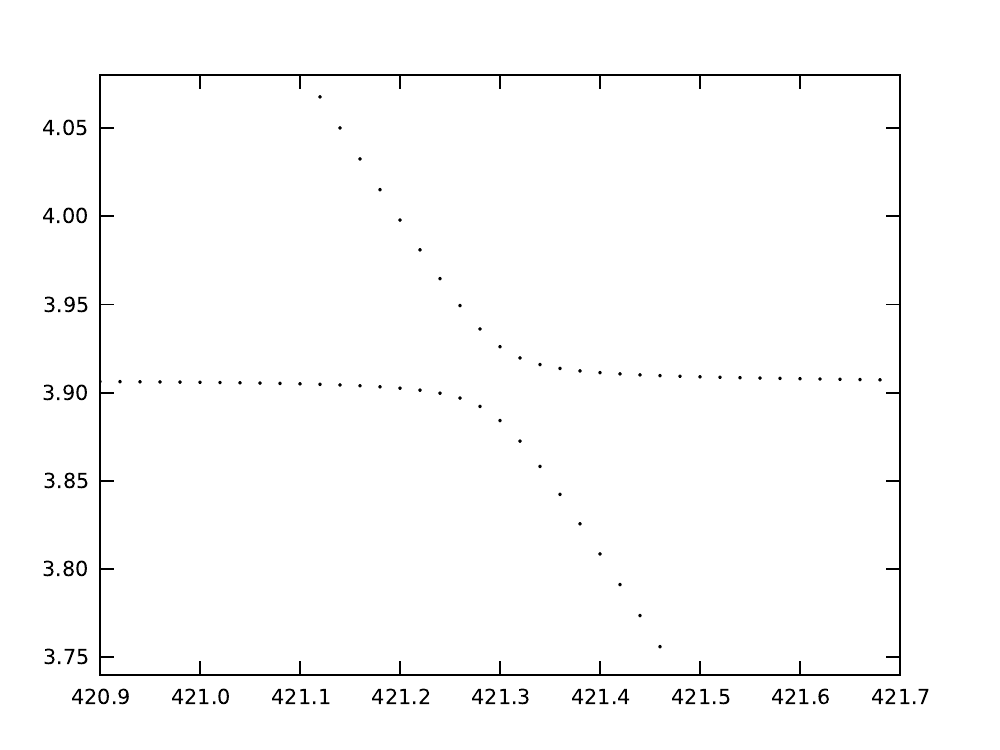}
\caption{In the left figure we have zoomed in on the time interval $[370,470]$ in Figure \ref{eigenblob}. 
In the right figure we have blown up the small rectangle in the left figure to observe more closely a crossing between the two largest eigenvalues.
The sampling rate was increased 50 fold to obtain the right figure, however the time scales between the two figures have been kept in sync to avoid confusion.}
\label{eigenzoom}
\end{figure}

Before moving on to the thermodynamics, we would like to point out something about the dynamics of the eigenvalues.
We can move the Vandermonde determinant into the exponential 
\begin{equation}
\mathcal{Z}_{\lambda} = \int d^N\lambda\, \delta(\tr(\Sigma_i \lambda_i))\exp\left( -\frac{\beta}{2}\sum_{i=1}^N \lambda_i^2 + 2\sum_{i<j}\log|\lambda_i - \lambda_j|\right)
\end{equation}
The exponential describes a quadratic potential with a logarithmic term.
Thus the eigenvalues should always repel each other and do not cross.
Although the rogue eigenvalue in Fig. \ref{eigenblob} appears to pass through the others at early times, it actually just transfers energy to the adjacent eigenvalue, like a Newton's cradle.
This behavior can be realized in plots like Fig. \ref{eigenblob} with a sufficiently small time step, a large enough sampling rate, and enough zooming.
An example of such is shown in Fig. \ref{eigenzoom}.

\begin{figure}[t]
\includegraphics{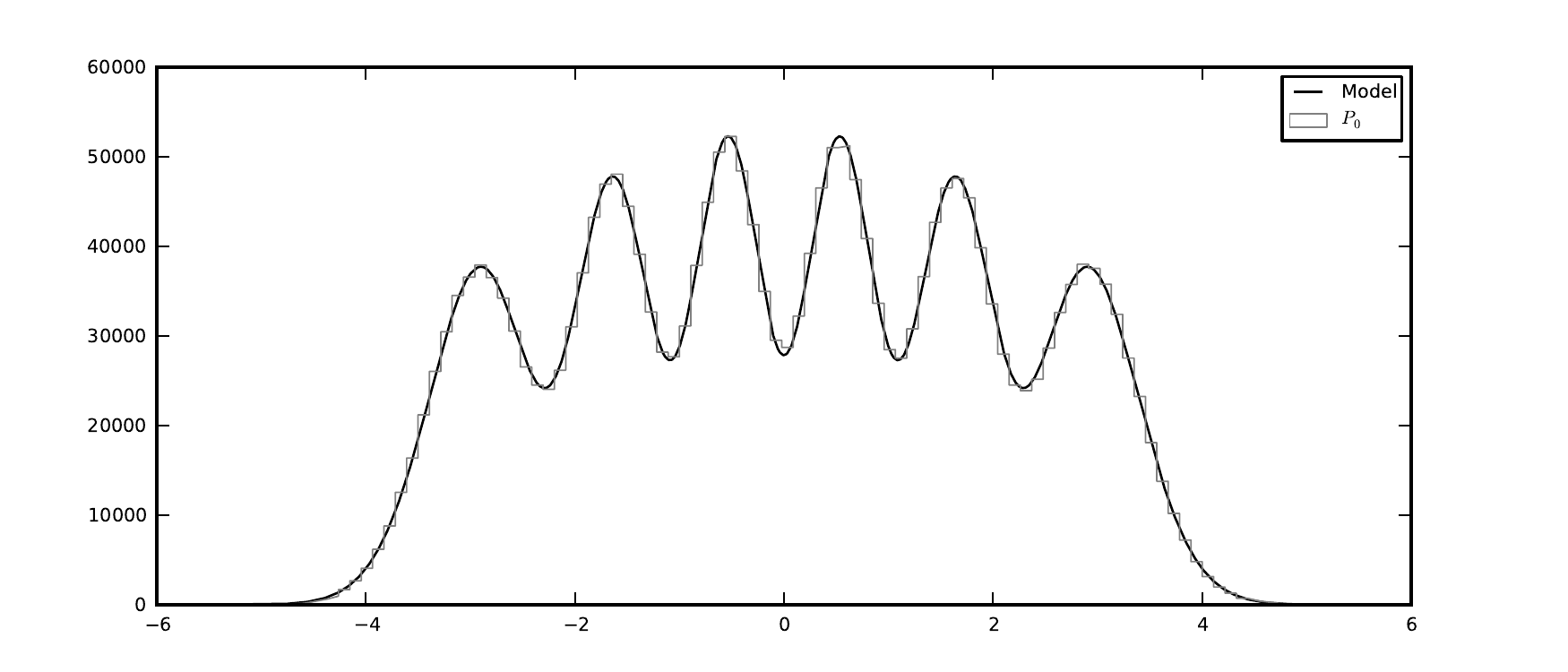}
\caption{A histogram of the eigenvalues of $P^0$ for a simulation of rank 6 matrices sampled after thermalization.
A total of $4\times10^5$ configurations were sampled meaning we fitted using $2.4\times 10^6$ eigenvalues.
By computing the second moment, we can fit the distribution to the level density of the TGUE.
The level density has been normalized to the total number of samples times the bin size.
The fit has an $R^2$ value of $0.999904$.}
\label{eigenfitp}
\end{figure}

To determine if our system has thermalized, the first step is to match our eigenvalue distribution to the 
distribution predicted by the partition function $\mathcal{Z}_{\lambda}$.
The Vandermonde determinant becomes exponentially complex with increasing $N$ and so a direct comparison is time expensive.
Instead we use the probability distribution obtained by integrating the partition function over all but one eigenvalue, i.e., the eigenvalue probability density or level density function.
An explicit form of the level density for the TGUE for arbitrary $N$ has recently been found in \cite{Ho-Kahn:2011}.
Fig. \ref{eigenfitp} shows that the eigenvalues of the traceless momentum matrices sampled over time after thermalization do indeed fit the predicted function.

In order to make these comparisons, we need to know the temperature $\beta$.
If we consider the tracelessness of the matrices as a constraint, then we can apply the proof in Appendix \ref{sec:temp} to a single traceless matrix and obtain
\begin{equation}
\label{constraintlesstemp}
T = \frac{\langle \tr(P^2)\rangle_0}{N^2 - 1}
\end{equation}
where $P$ is any single momentum matrix and the zero subscript indicates we take the expectation value with respect to the TGUE.
For numerical measurements, the zero subscript indicates averaging only over the traceless part of the matrices.
The momentum contribution to the Hamiltonian is invariant under an $\text{SO}(9)$ transformation and one would suspect that we could use any momentum matrix and obtain the same temperature.
The numerical data of Table \ref{isotropyfail} shows that this is not the case.

What we have forgotten about in developing our thermometer are the constraints. 
Indeed, if we read directly from the table under our naive assumptions, we would conclude that the system has two different temperatures and therefore has not thermalized, even though it has isotropized along the relevant directions.
It would seem that either the system is not thermal or the thermometers are broken.
To answer this puzzle notice that the naive Gibbs ensemble is over all $P$, $Q$ matrices, but our true ensemble is over $P$, $Q$, $X$, $Y$ matrices, subject to the gauge constraint.
The gauge constraint can not be ignored: it is an essential part of the dynamics. 
For example, if we ignored it we would get the wrong specific heat for the system. 
In quantum systems this is usually cured by adding ghosts, but in this case we need to add the constraint to the ensemble. 

The gauge constraint induces symmetry breaking between the $P$, $Q$ variables which breaks the $\text{SO}(9)$ symmetry, but preserves the $\text{SO}(3)\times \text{SO}(6)$ subgroup.
The $\text{SO}(9)$ symmetry is broken between the $X$ and $Y$ matrices by the dynamics and thus the typical values of the $X$ and $Y$ matrices are different.
This breaking of $\text{SO}(9)$ therefore induces an asymmetry in the Gibbs ensemble for the $P$, $Q$ matrices because they appear with different coefficients in the constraint.
We also have conservation of angular momentum in the $X-P$ and $Y-Q$ planes, so we should also remove the angular momentum degree of freedom as a constraint: we chose initial configurations with zero angular momentum. 
Thus the $\text{SO}(3)\times \text{SO}(6)$ subgroup is restored on averaging throughout the entire simulation and explains why in Table \ref{isotropyfail} it appears that each of the $P_i$ yield the same temperature as do each of the $Q_a$, but independently they seem to be different.
\begin{table}[b] 
\begin{tabular}{|c||c|c|c|c|c|c|c|c|c|}
\hline
$N$ & $\langle\tr(P_0^2)\rangle_0$ & $\langle\tr(P_1^2)\rangle_0$ & $\langle\tr(P_2^2)\rangle_0$ & $\langle\tr(Q_1^2)\rangle_0$ & $\langle\tr(Q_2^2)\rangle_0$ & $\langle\tr(Q_3^2)\rangle_0$ & $\langle\tr(Q_4^2)\rangle_0$ & $\langle\tr(Q_5^2)\rangle_0$ & $\langle\tr(Q_6^2)\rangle_0$ \\
\hline
4  & $23.2 \pm 0.6$ & $23.3 \pm 0.4$ & $23.2 \pm 0.5$ & $21.3 \pm 0.5$ & $21.3 \pm 0.5$ & $21.2 \pm 0.6$ & $21.2 \pm 0.4$ & $21.3 \pm 0.4$ & $21.0 \pm 0.4$ \\
11 & $26.9 \pm 0.3$ & $27.2 \pm 0.2$ & $27.0 \pm 0.3$ & $26.6 \pm 0.2$ & $26.5 \pm 0.3$ & $26.6 \pm 0.2$ & $26.6 \pm 0.3$ & $26.6 \pm 0.2$ & $26.5 \pm 0.2$ \\
23 & $32.2 \pm 0.3$ & $32.2 \pm 0.2$ & $32.1 \pm 0.2$ & $31.9 \pm 0.2$ & $31.9 \pm 0.2$ & $31.9 \pm 0.2$ & $31.9 \pm 0.2$ & $31.9 \pm 0.2$ & $32.0 \pm 0.2$ \\
\hline
\end{tabular}
\caption{Time average samples of the trace of the square of the traceless momenta for various $N$ with $v = 20.0$, $\hbar = 0.001$, and $\delta t = 0.001$.
For each expectation value, 20 samples were used but the number of configurations varies with $N$ due to limited hard disk space.
The differences between the values in the disjoint groups of $\langle\tr(P_i^2)\rangle_0$ and $\langle\tr(Q_a^2)\rangle_0$ are smaller compared to the differences of the values in the union of these two groups indicating the breaking of the $\text{SO}(9)$ symmetry between the momenta that is present the Hamiltonian.}
\label{isotropyfail}
\end{table}

Putting everything together, our true Gibbs distribution is the GUE for nine matrices with the constraint of tracelessness, the gauge constraint, and the total amount of angular momentum set to zero.
Each of these constraints is linear in the momenta and thus we may apply the result in Appendix \ref{sec:temp}, 
which is a generalized equipartition theorem, to obtain an absolute normalization of the temperature.
\begin{equation}
\label{calibratedtemp}
\sum_{i=0}^2\langle \tr(P_i^2)\rangle_0 + \sum_{a=1}^6\langle \tr(Q_a^2)\rangle_0 = \left( 9(N^2 - 1) - (N^2 - 1) - \frac{3 \cdot 2}{2} - \frac{6\cdot 5}{2}\right) T = (8N^2 - 26)T.
\end{equation}
The lowest rank matrix which can exhibit thermalization behavior nontrivially is $N = 2$ and so we do not run into an issue of negative temperature (this would indicate that we have not taken into account relations between the constraints).

We would also like to ensure that temperature measurements are independent of the time step parameters in our numerical implementation, that is, $\delta t$.
The BMN matrix model exhibits chaos and so shrinking the time step does not cause the numerical solution to converge to a solution of the equations of motion as small differences grow exponentially for large times.
We still expect that each of the trajectories computed this way would lead to the same ensemble since we should be sampling the phase space according to the dynamically invariant measure, in a manner typical of numerical simulations of chaotic dynamical systems.
\begin{table}[t]
\begin{tabular}{rl|>{\rule{-2pt}{0pt}}c|c||c||c||}
\cline{3-6}
 & & & \multicolumn{3}{c||}{$N$} \\
\hline
\multicolumn{2}{|c|}{$\delta t$} & & $4$ & $14$ & $23$ \\
\hline
\multicolumn{1}{|r@{.\hspace{-1.5pt}}}{0} & 005    & & $1.932 \pm 0.014$ & $0.16235 \pm 0.00050$ & $0.06845 \pm 0.00018$ \\
\multicolumn{1}{|r@{.\hspace{-1.5pt}}}{0} & 003125 & & $1.932 \pm 0.014$ & $0.16234 \pm 0.00043$ & $0.06844 \pm 0.00017$ \\
\multicolumn{1}{|r@{.\hspace{-1.5pt}}}{0} & 0025   & & $1.932 \pm 0.014$ & $0.16235 \pm 0.00047$ & $0.06843 \pm 0.00016$ \\
\multicolumn{1}{|r@{.\hspace{-1.5pt}}}{0} & 002    & & $1.932 \pm 0.016$ & $0.16234 \pm 0.00048$ & $0.06843 \pm 0.00018$ \\
\multicolumn{1}{|r@{.\hspace{-1.5pt}}}{0} & 00125  & & $1.932 \pm 0.013$ & $0.16234 \pm 0.00050$ & $0.06843 \pm 0.00016$ \\
\multicolumn{1}{|r@{.\hspace{-1.5pt}}}{0} & 001    & & $1.932 \pm 0.014$ & $0.16234 \pm 0.00048$ & $0.06843 \pm 0.00016$ \\
\multicolumn{1}{|r@{.\hspace{-1.5pt}}}{0} & 000625 & & $1.932 \pm 0.015$ & $0.16234 \pm 0.00053$ & $0.06843 \pm 0.00016$ \\
\multicolumn{1}{|r@{.\hspace{-1.5pt}}}{0} & 0005   & & $1.932 \pm 0.017$ & $0.16234 \pm 0.00051$ & $0.06843 \pm 0.00017$ \\
\multicolumn{1}{|r@{.\hspace{-1.5pt}}}{0} & 0004   & & $1.932 \pm 0.015$ & $0.16234 \pm 0.00051$ & $0.06843 \pm 0.00015$ \\
\hline
\end{tabular}
\caption{Measured temperatures of thermalized system for rank $4$, $14$, and $23$ matrices for several $\delta t$ using the same initial conditions for each $N$.
The sampling rate is chosen such that the time separation between recorded configurations is kept constant, in particular $(\text{sampling rate}) \times (\delta t) = 0.05$.
A total of 20 samples were used for each measurement, however, the number of configurations per sample was decreased with increasing $N$.}
\label{dtindependence}
\end{table}

In order to measure the temperature we need to measure $\langle \tr(P^2)\rangle_0$ for all momenta matrices.
We can determine the accuracy of our measurements by dividing the configurations into groups to obtain several sample measurements of $\langle \tr(P^2)\rangle_0$.
The expectation value is independent of the grouping of configurations due to its linearity, but now we can obtain a standard deviation.
Consecutive configurations are correlated, so some care must be taken when grouping configurations to make a sample.
To minimize correlations between samples, we group configurations consecutively.
Each sample will be correlated with itself, but different samples will only be correlated at their boundaries.
This sampling process provides a way to measure $\langle \tr(P^2)\rangle_0$ with some degree of accuracy.
The temperature is computed using equation (\ref{calibratedtemp}) and the standard deviation is computed by summing the standard deviations of the $\langle \tr(P^2)\rangle_0$ in quadrature.

Table \ref{dtindependence} lists the temperatures of simulations for various $N$ and various $\delta t$ with the same initial conditions for each $N$.
The temperatures are equal to several significant figures and the error bars intersect a common average value.
Furthermore we find that the coefficients of variation are less than $1\%$.
To claim our comparison among different $\delta t$ is reasonable, the sampling rate is chosen such that the time between recorded configurations is constant.
We conclude that the temperature is a well defined quantity regardless of how far apart trajectories in phase space become due to changing the time step. 

Some simple observables that are also natural thermodynamic variables are the sizes of the distributions of $X$ and $Y$ matrix eigenvalues.
We can determine how they scale with the temperature and $N$ using the virial theorem. 
For our case, a virial can be computed for each $X$ and $Y$ matrix. 
Consider, for simplicity, the expression
\begin{equation}
\frac{d} {dt} \tr( X^i P_i) = \tr (P_i^2) + \tr( X^i \dot P_i) = \tr(P_i^2) - \tr( X^i \partial_{X^i} V), 
\end{equation}
where we are only using the traceless parts of the matrices (we subtract the trace modes, which are decoupled) and we don't sum over $i$. 
We then integrate over a period of time $\tau$ and average and obtain
\begin{equation}
\frac{1}{\tau} (\tr(X^i P_i)(\tau) - \tr(X^i P_i)(0)) = \left\langle \tr(P_i^2) - \tr(X^i \partial_{X^i} V) \right\rangle_\tau
\end{equation}
If the trajectories are bounded then the left hand side asymptotes to zero as $\tau\to \infty$ and we obtain a relation between the kinetic energy and various derivatives of the potential energy.
This is simplest in the BFSS matrix model.
We find after summing over the $X_i$ that  
\begin{equation}
\sum_i \tr(P_i^2) + \sum_{ij} \tr([X^i,X^j][X^i,X^j]) = 0.
\end{equation}
This is, we find that the kinetic energy is twice the potential energy $2 E_{\text{kin}} = 4 E_{\text{pot}}$. We have already argued that the left hand side grows like $(8 N^2-26) T$, so we find that the right hand side takes the same value.
The total energy in that case is
\begin{equation}
E_{tot} = \frac{3}{4}(8N^2 - 26)T
\end{equation}
At large $N$, we get that the energy as a function of the temperature is $\frac{3}{4}(8N^2)T$.
The specific heat is essentially the same as that of $6N^2$ harmonic oscillators and is constant.
Notice that this result also matches the Monte-Carlo lattice simulations in the matrix model, as seen in Fig. $3$ of \cite{Catterall:2008yz}.
Other such simulations \cite{Hanada:2008ez} do not cover the high temperature regime.

\begin{figure}[h]
\includegraphics{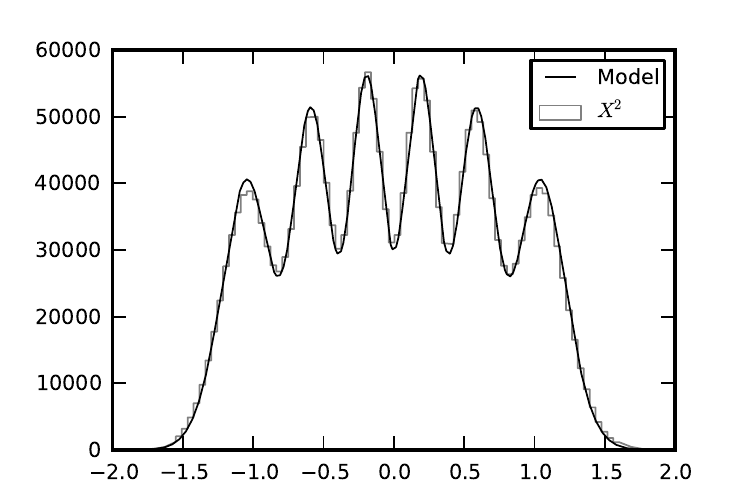}
\includegraphics{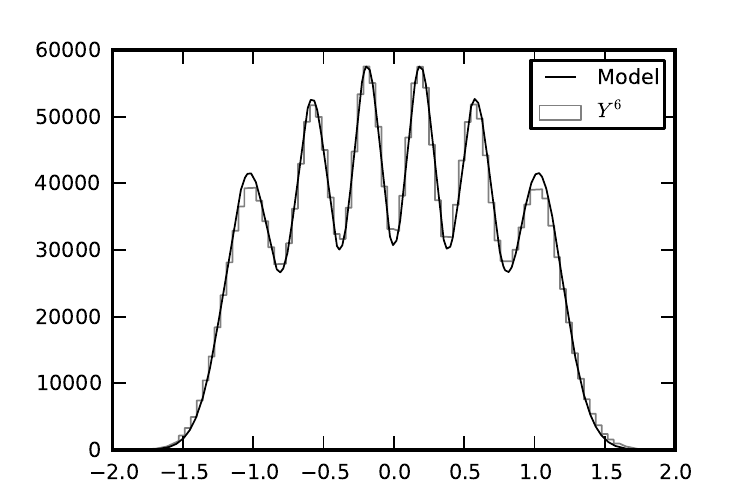}
\caption{A histogram of the eigenvalues of $X^2$ and $Y^6$ for a simulation of rank 6 matrices sampled after thermalization.
A total of $4\times10^5$ configurations were sampled meaning we fitted using $2.4\times 10^6$ eigenvalues.
Each fit has an $R^2$ value of $0.99925$ and $0.9988$ respectively.}
\label{eigenfitx}
\end{figure}

Another means of getting at the size of the $X$ and $Y$ matrices is to look at the distribution of the eigenvalues.
The elements of any single coordinate matrix appear at most quadratically in the Hamiltonian.
Integrating the Gibbs distribution $\exp(-\beta H)$ over all momenta and all but one coordinate matrix will be a Gaussian distribution in the remaining coordinate matrix elements.
The standard deviation will be modified due to the constraints, but the form of the integrand will remain unchanged.
The only constraint left over is the tracelessness of the matrices.
Thus we expect that the eigenvalues follow the level density of the TGUE.
Observing Fig. \ref{eigenfitx} this is exactly what we see.

\begin{figure}[h]
\includegraphics{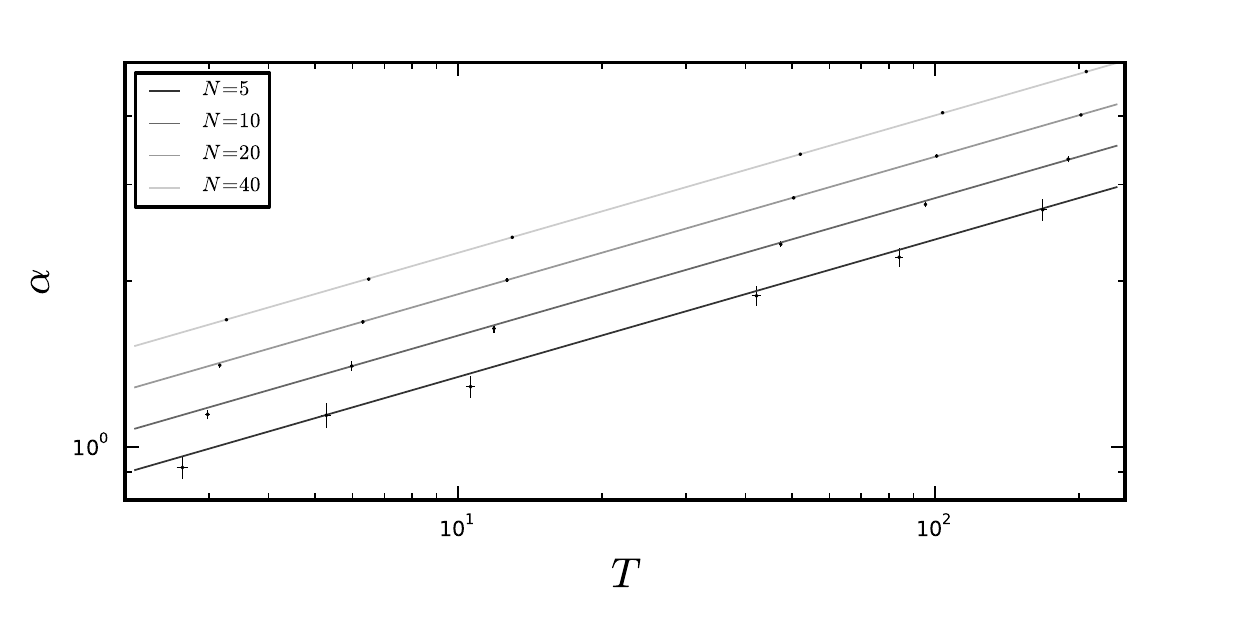}
\caption{A plot of the size of $X^0$, measured as $\alpha = \sqrt{\langle \tr(X^2)\rangle / N}$, as a function of temperature for various $N$.
The error bars for larger $N$ are smaller than the point size and thus may not be visible in the plot.
The lines are given by $\alpha = cN^{1/4}T^{1/4}$ for constant $N$.
Doing a least squares fit gives the constant $c = 0.504$.
Plots for the other coordinate matrices give identical results.}
\label{scalingfit}
\end{figure}

We can also estimate the commutator squared term if we assume that the $X$ and $Y$ matrices are random.
If the eigenvalues of $X$ are of order $\alpha$ (we call this the size of the matrix), then the eigenvalues of $[X^i,X^j][X^i,X^j]$ grow like $\alpha^4$, and we get that $N \alpha^4 \simeq N^2 T$.
Thus the size of the matrices grows like $\alpha \simeq N^{1/4} T^{1/4}$.
This is also true for the BMN matrix model at high temperature. 
In that case, the cubic and quadratic terms in the potential are subleading when the size of the matrices gets large and one asymptotically matches the BFSS matrix model.
A test of our prediction is shown in Fig. \ref{scalingfit} with remarkable agreement.
Our claims hold for large $N$ and so we see larger deviations for smaller $N$.

%
%

\section{Power spectra and classical chaos}
\label{sec:chaos}

As we have seen, there is evidence for thermalization in the BMN matrix model.
Similar considerations show that the BFSS matrix model thermalizes (this has been studied for different initial conditions in \cite{Riggins:2012qt}).
This should not be surprising. 
Both the BFSS and BMN matrix models result from dimensional reduction of SYM to constant configurations on either flat space or a sphere.
It turns out that the dynamics of translation invariant configurations of Yang Mills theories generally exhibit chaos \cite{Sav, Chirikov:1981cm} and therefore the BFSS matrix model exhibits chaos (this was reiterated in \cite{Aref'eva:1997es}).
Because of classical scale invariance of the Yang Mills action, this extends all the way to infinitesimal configurations of the fields.
Chaos is also present if a mass term is added \cite{Matinyan:1981ys}, but to access the chaotic region requires finite field configurations.
The BMN matrix model is effectively a massive version of the BFSS matrix model, so it should also exhibit chaos for field configurations where the fields are sufficiently large, but can display integrable behavior for small oscillations around vacuum states. 
 
In this section we analyze the chaos in both the BFSS and BMN matrix models 
and show how we can use this information to study holography. 
For this purpose, let us pretend that a configuration in our classical system represents a thermal equilibrium state in a quantum system.
Then we would be interested in various response functions and correlation functions of observables in order to understand the dynamics of the thermal state.
 
For example, a typical gauge invariant observable would be a trace. 
The simplest traces are those of the matrices $X^i$ and $P_i$.
However, we don't gain much from studying these as they are decoupled and either work as a harmonic oscillator (this is the center of mass motion in the BMN matrix model), or they give free non-relativistic motion on flat space (this is the center of mass motion in the BFSS matrix model). 
So we need to look for traces of more complicated composite objects.
Here it pays to notice that we are studying configurations with zero angular momentum. 
For black holes, this means we are looking for spherically symmetric configurations in ten dimensions, so the perturbation modes of the black holes may be characterized by their angular momenta. 
The matrices $X^i$ form a $\mathbf{9}$ of $\text{SO}(9)$ in the BFSS matrix model (in the BMN matrix model the appropriate group is $\text{SO}(6)\times \text{SO}(3)$, which is only slightly more complicated to analyze), so it is convenient to study configurations that are highest weight states of $\text{SO}(9)$ multiplets.
Spherical symmetry then predicts that the one point correlation functions of $\text{SO}(9)$ non-singlets in a thermal state would be zero, but two point functions could be non-zero if there is a singlet in the tensor product of the two $\text{SO}(9)$ representations.
 
Let $Z= X^1+iX^2$. This is a highest weight state of $\text{SO}(9)$, 
so we can also take operators $\CO_L=\tr(Z^L)$. 
These will  be highest weight states of symmetric traceless combinations of the $X$ with angular momentum $L$. 
In the case of $\mathcal{N}=4$ SYM theory such modes constructed from scalars at zero temperature are protected states 
\cite{Corley:2001zk} and they are the dual description of gravity excitations of AdS space. 
These states already display incompressible and dissipation free  hydrodynamic behavior, at zero temperature, 
in the form of a quantum hall droplet \cite{Berenstein:2004kk}. 
Collective excited states can be put in one to one correspondence with gravity states \cite{Lin:2004nb}
and the shape of the gravity configuration is directly determined by the expectation values of these traces. 
We expect that these simplest traces are also closely related to gravity modes of a black hole in the
in the dual of the 
$\mathcal{N}=4$ SYM theory at finite temperature.
This leads us to expect that these dynamical variables are also 
related to gravity modes in both the BFSS and BMN matrix models at finite temperature. 
For example, they could describe how gravitational or dilaton partial waves (see, e.g., \cite{Das:1996wn, Klebanov:1997kc}) 
are absorbed or emitted from such systems.
In any case, these variables are important for understanding these configurations 
in detail and could help us to ultimately learn about emergent black hole phenomena like Hawking radiation, 
or whatever corresponds to it in the regime we are studying.

The conjugate operators are $\bar \CO_L = \tr(\bar Z ^L)$. An interesting two point function is then given by
\begin{equation}
\label{eq: corr}
\vev{ \CO_L(t) \bar \CO_L(t')} = S_L(t-t'),
\end{equation}
which, owing to time translation invariance of the thermal density matrix and the Hamiltonian, can only depend on the combination $(t-t')$. One can also study this correlation in Fourier space, so that $\tilde S_L(\omega)= \int S_L(a) \exp(-i \omega a) da$, which is how frequency dependent transport coefficients are usually defined. 
Closely related quantities can be calculated in gravitational setups by using the holographic dictionary 
in a perturbed black hole geometry with infalling boundary conditions at the horizon. 
This ultimately gives a relation between the quasinormal modes of asymptotically AdS black holes
and CFT response functions (see \cite{Kokkotas:1999bd, Berti:2009kk, Konoplya:2011qq} for reviews).
We need to understand how we can compute these quantities in the classical dynamics. 
The key observation is that in quantum chaotic regimes we expect $S_L(t - t')$ to be roughly independent of the microstate that we choose, even if it is a pure state, so long as it is a typical state of the thermal system
\footnote{This expectation is an extension of the idea that all energy eigenstate behave as if they are thermal states for time independent questions \cite{Srednicki}.
Some evidence of this behavior can be found in examples \cite{MWH}.}. 
Indeed, we expect most correlation functions of energy eigenstates $|E_i\rangle$ to be approximately thermal: 
\begin{equation}
\vev{ E_i | \CO_L(t) \bar \CO_L(t')| E_i} \simeq S_L(t - t'),
\label{eq: correigen}
\end{equation}
and the dependence on $t - t'$ is guaranteed by time translation invariance of the matrix elements of the energy eigenstates. 
Here thermalization means that $S$ decays to zero (usually exponentially), with some thermalization time $\tau$, and that the left hand side approximates the right hand side on time scales that are short compared to Poincar\'e recurrence times.

We extend this to more typical states, which are superpositions of energy eigenstates around some energy value, 
by averaging over $t$ keeping $t-t'$ fixed. We then expect
\begin{equation}
\left\langle \vev{ \psi| \CO_L(t) \bar \CO_L(t+a) |\psi} \right\rangle_t  \simeq S_L(a),
\label{eq: corrtyp}
\end{equation}
where $|\psi\rangle$ is some typical state and and we average over time. 
The quantum theory should match the classical theory for these correlation functions in the correspondence limit: 
at very large quantum numbers the answer in the classical and quantum theory should be very similar so long as the time scales involved are relatively small compared to the Poincar\'e recurrence time. 
For a many body system like our large $N$ matrix models, all the time scales we consider will be much smaller 
than the recurrence time. 
The is an example of using classical statistical mechanics as an approximation to quantum statistical mechanics.

\begin{figure}[ht]
\includegraphics[width=4 in]{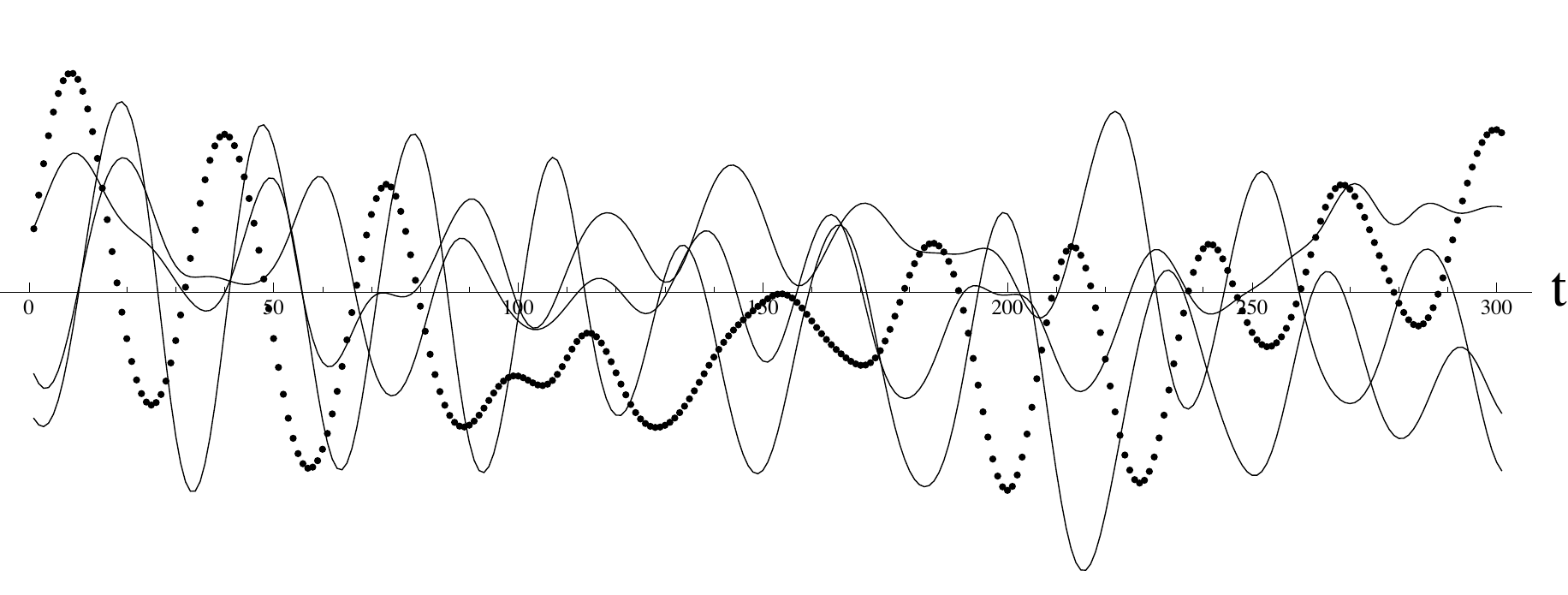}
\caption{Time series of the function $\Real(\tr(Z^2(t)))$. We show four different time series obtained by rotations of the $Z$ by $\text{SO}(9)$ action.
They all look similar, showing approximate rotation invariance of the time average.
We showcase the discrete data we have in one of them, so show that the time dependent features are well covered by our time slicing.
The sample shown here is for $18\times 18$ matrices.}
\label{fig:timeseries}
 \end{figure}

We compute the left hand side of \eqref{eq: corrtyp} by using a typical state of the microcanonical ensemble and averaging over its trajectory. 
We obtain these from our simulations by first waiting until the system thermalizes then averaging over various configurations. The functions $S_L(a)$ are the autocorrelation functions of the system. 
Let us first consider the time series of $\CO_L(t)$ for some $L$ after thermalization. We display this in Fig. \ref{fig:timeseries}. 
From the figure we see oscillations of a typical frequency, but they are not regular nor centered on zero. 
Rather, they appear to be superposed on waves of a much longer period than the time period shown.

\begin{figure}[ht]
\includegraphics[width=3.5 in]{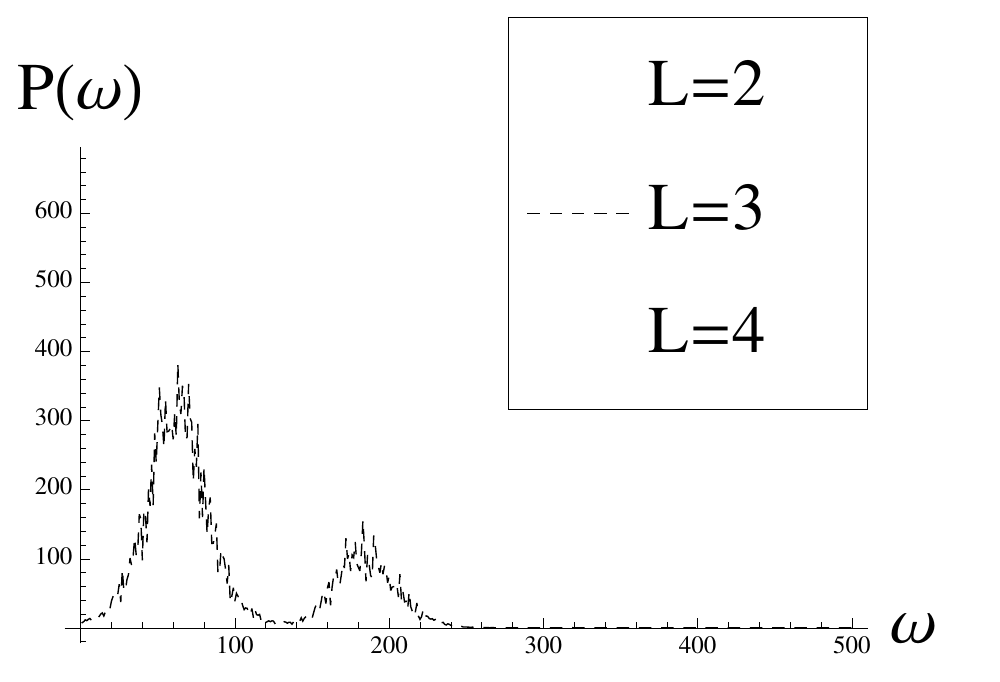}
\caption{The power spectrum of $S(a)=\langle \tr(X^1 + iX^2)^L(t) \tr(X^1 - iX^2)^L(t+a)\rangle$ for various $L$ in random units.
Results shown are for $13 \times 13 $ matrices in the BFSS matrix model after thermalization.
The results are averaged over 15 runs of the same length, taken from splitting a single time series in $15$ equal parts.
The jiggling of the data should be interpreted as an estimate of the statistical error bars for each frequency.}
\label{fig:powspect}
\end{figure}

To extract information from the time series, we compute $\tilde S_L(\omega)$ by taking the Fourier transform of 
$\CO_L(t)$ and averaging over many configurations.
In our case, rotation invariance tells us that we did nothing special by choosing $(X^1+i X^2)$ as our highest weight state.
Since all tensor representations of $\text{SO}(n)$ are real $(X^1 - iX^2)$ can be obtained from a rotation of $(X^1 + iX^2)$.
This means that the right hand side of \eqref{eq: corrtyp} only depends on the absolute value of $a$ and therefore the power spectrum $P(\omega) = \tilde{S}(|\omega|)$ is an even real function of $\omega$.
Thus we only have to display the answer for $\omega\geq 0$. 
We show power spectra for $L=2$, $3$, $4$ in Fig. \ref{fig:powspect}.

The first observation we can make from these plots is that the power spectra are those of a chaotic system.
If a system is integrable, we expect the system to be solvable in terms of action-angle variables.
The angle variables are multivalued, with period $2\pi$ or $1$ depending on conventions, 
but they have very simple time dependence $\phi_\alpha(t)=\phi_\alpha(0)+\omega_\alpha t$.
Any single valued function on phase space can then be represented by its Fourier series in the angle variables 
and its time dependence is that of a quasi-periodic function of time, 
with characteristic frequencies determined by all integer linear combinations of the $\omega_\alpha$.
Thus, the power spectrum of its time series should display delta-function peaks at the characteristic frequencies of the system.
The series we observe has no delta function peaks, rather it seems to be described by broad band noise. This is one of the standard criteria to distinguish chaotic from non-chaotic systems \cite{Eckmann:1985zz}.

\begin{figure}[ht]
\includegraphics[width=3.5 in]{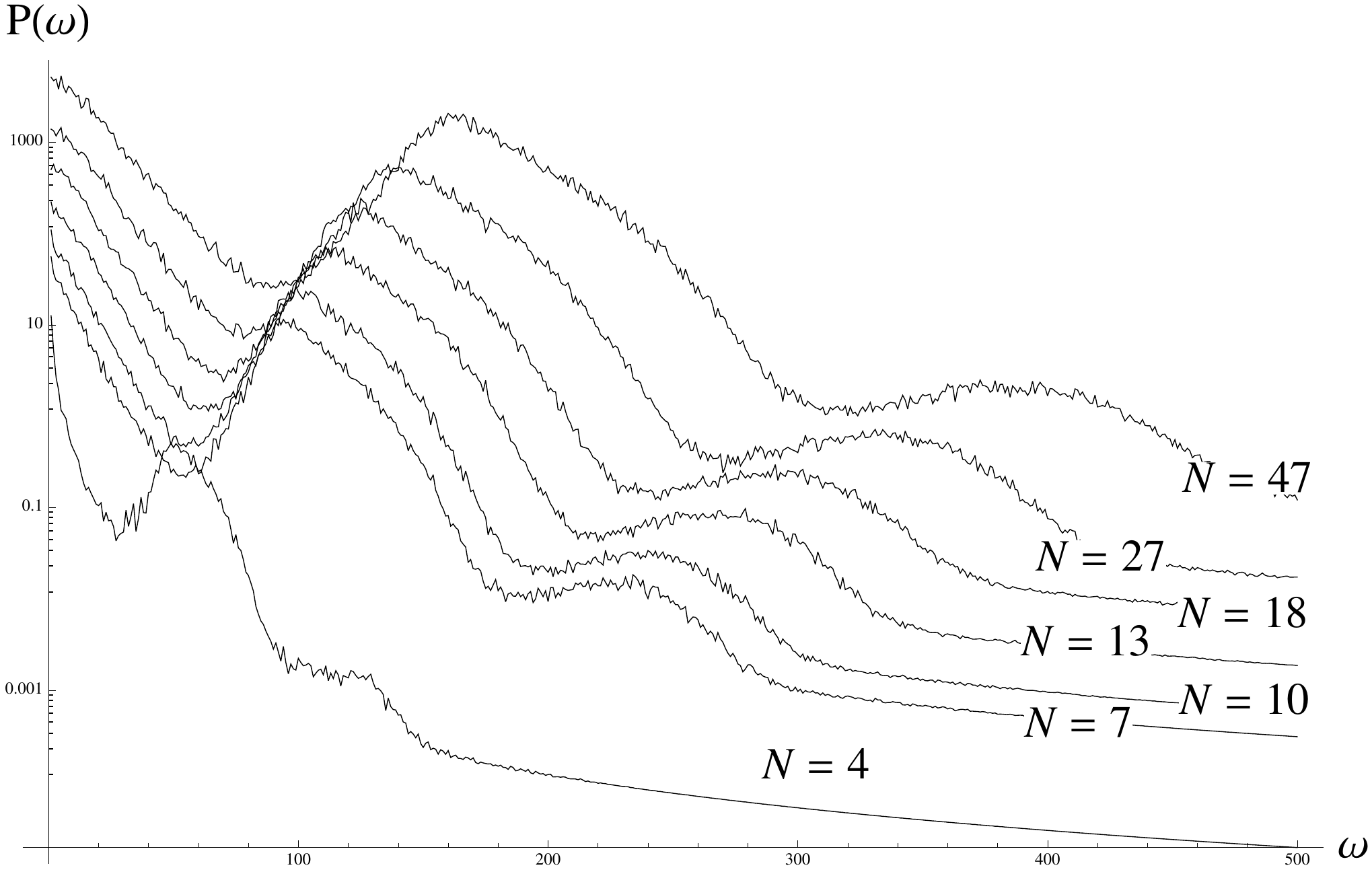}
\caption{The power spectrum of $ \tr(X^1+i X^2)^2(t)$ for various sizes of $N\times N$ matrices in random units (its Fourier transform is $S(a)$).
Results shown are for the BFSS  matrix model after thermalization.
The results are averaged over 15 runs of the same length, taken from splitting a single time series in $15$ equal parts.
We also average further over $8$ rotations of the variables to increase the statistics.
The jiggling in the curves gives a measure of the statistical error bars of the data sets.}
\label{fig:spect_l2}
 \end{figure}

Now, we can also make more sense of what we see in Fig. \ref{fig:timeseries}. 
Notice that for $L=2$ there seem to be two peaks, one near zero and another one at a characteristic frequency $\omega_0$.
Thus we may describe the signal approximately as oscillating with some characteristic frequency 
while riding on a very low frequency envelope.
We also have information on other modes.
For $L=4$ we observe broader peaks located roughly in the same places as well as a frequency doubling of the $\omega_0$ peak.
For $L=3$ we observe peaks at `half period' spacings relative to $L=2$.
The reader may have noticed that in Fig. \ref{fig:timeseries} we used $18\times 18$ matrices, whereas in Fig. \ref{fig:powspect} we studied instead the system of $13\times 13$ matrices and may be worried.
The natural way to understand this is to look at how the power spectrum depends on $N$, the size of the matrices.
This is shown in Fig. \ref{fig:spect_l2}.

\begin{figure}[ht]
\includegraphics[width=3.5 in]{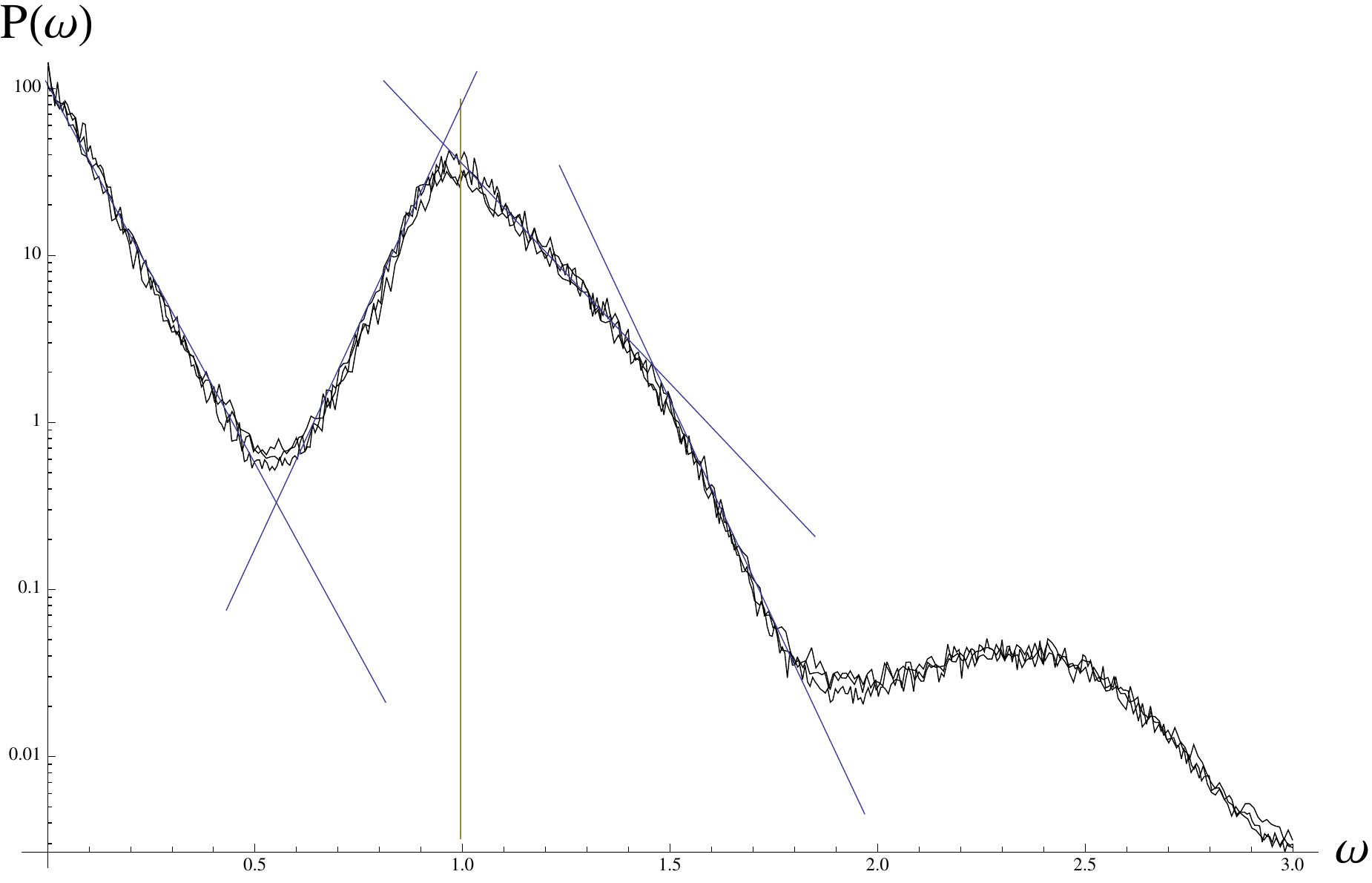}
\caption{The power spectrum of $ \tr(X^1+i X^2)^2(t) $ for various sizes of $N\times N$ matrices.
The axis of frequency has been rescaled for each $N$, to the frequency $\omega_N$, and we have also rescaled the power spectrum. The reference frequency for each $N$ is located  at $1$ in the graph.
Results shown for $N=7, 10, 47$.
We also have drawn additional suggestive straight lines superposed on the graph that serve as distinctive features of the power spectrum.}

\label{fig:collapse}
 \end{figure}

We see from Fig. \ref{fig:spect_l2} that the power spectrum of all $L=2$ modes for various values of $N$ are actually very similar to each other.
Each has a large peak at zero, which is more noticeable in a log-linear plot.
Also, we see a second peak at some characteristic frequency which depends on the energy of the system and $N$.
We compare various values of $N$ by finding the location of the peak and rescaling 
the power to make the plots lie on top of each other.
To find the location of the peak we do a fit of the $\log (P(\omega))$ 
to a quadratic function of $\omega$ in a small interval around the visual maximum.
We then extract the value of $\omega$ that corresponds to the maximum and we scale each axis of frequency to the corresponding $\omega_N$ found for each $N$.
The main systematic error comes from the choice of the interval.
We show the result in Fig. \ref{fig:collapse}.
The data have collapsed to a single graph. 

\begin{figure}[ht]
\includegraphics[width=3.5 in]{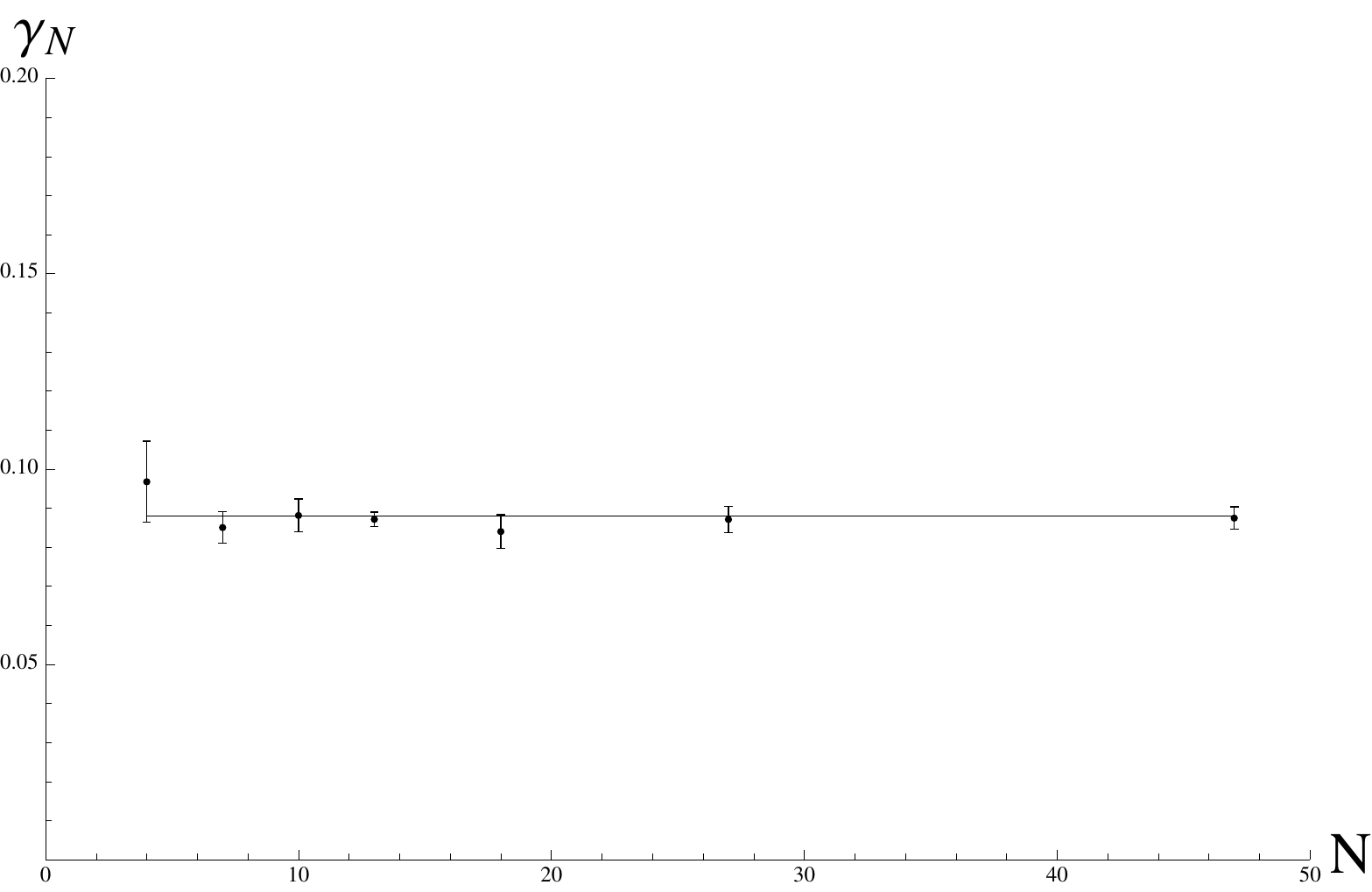}
\caption{$\gamma_N$ versus $N$.
The error bars indicate systematic errors from choosing the fitting intervals. 
A fit to a single number has been done ignoring $N=4$ which is an outlier by inspection.
All of the different values of $N>4$ align within the systematic errors.}
\label{fig:gamma}
\end{figure}

As shown suggestively in Fig. \ref{fig:collapse}, the logarithm of the power spectrum seems to have rather distinct 
features that are characterized by straight lines.
We can numerically compare the different values of $N$ to get an idea of how closely the curves match 
by considering the width of the peak near zero, relative to $\omega_N$.
The dimensionless width can be parametrized by the slope 
\begin{equation}
\gamma_N = -\left( \frac{d \log (P(\omega)_N)}{d\omega} \omega_N\right)^{-1}, 
\end{equation}
which is evaluated near zero and with a cutoff slightly below $0.4 \omega_N$. 
Larger $\gamma_N$ corresponds to a larger width.
This is a dimensionless number that can be used to quantify how close the curves at different $N$ are to each other.
We show this in Fig. \ref{fig:gamma}.
As seen in the figure, all values of $N > 4$ have close to same behavior and the differences are controlled by the 
systematics of the fit, which is dominated by the choice of interval over which we compute the slope.
This matching is necessary to have a well defined large $N$ limit for these time dependent correlation functions.
We have checked that the graphs are very similar for other simple operators and 
so they all seem to have a good large $N$ limit.

\begin{figure}[ht]
\includegraphics[width=3.5 in]{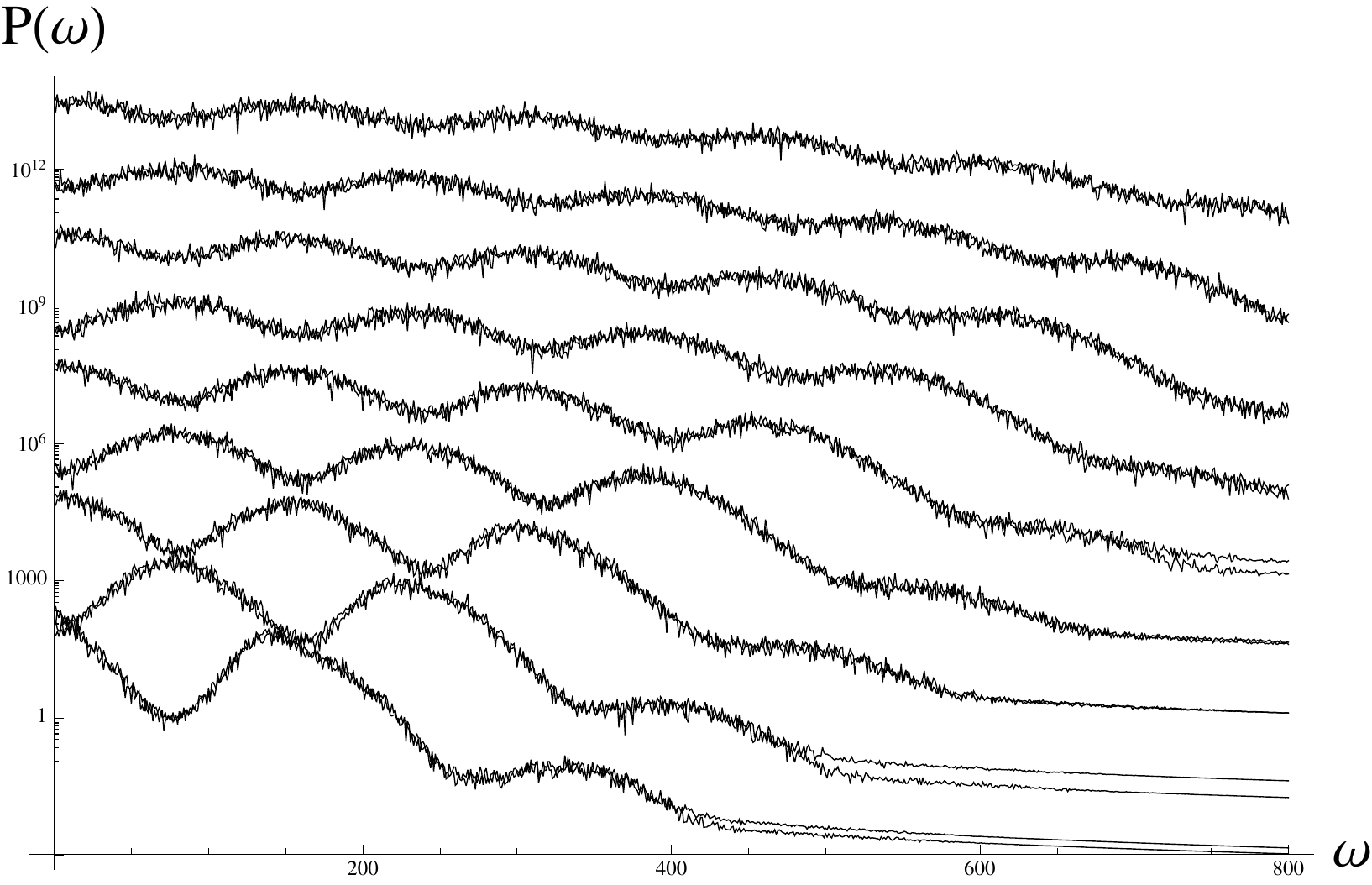}
\caption{Power spectrum in arbitrary units for $\CO_L$, with $L=2, \dots 10$, with values of $L$ increasing from bottom to top in the graph. The plots are vertically separated so they can be easily distinguished.
For each $L$ we show two such sets.
These data are from $N = 27$.}
\label{fig:pow27}
\end{figure}

How should we interpret these results?
One way is to conclude that the system is behaving hydrodynamically: 
there are some large $N$ collective variables whose time dependent characteristics are independent of $N$, 
up to some rescalings of the variables by the natural frequency of the dynamics.
We just checked the simplest angular momentum mode with $L=2$, but we can do the analysis for $\tr(Z^L)$ for various $L$.
The plot of the power spectrum for various $L$ can be seen in Fig. \ref{fig:pow27}.
As the reader can see, the patterns observed for low $L$ in Fig. \ref{fig:powspect} persist.
Notice that the logarithmic scaling of the power spectrum makes the pattern more regular.
To investigate this in more detail we need to address further how the different $N$ are related to each other.
We will do this in the next section. 

\begin{figure}[ht]
\includegraphics[width=3.5 in]{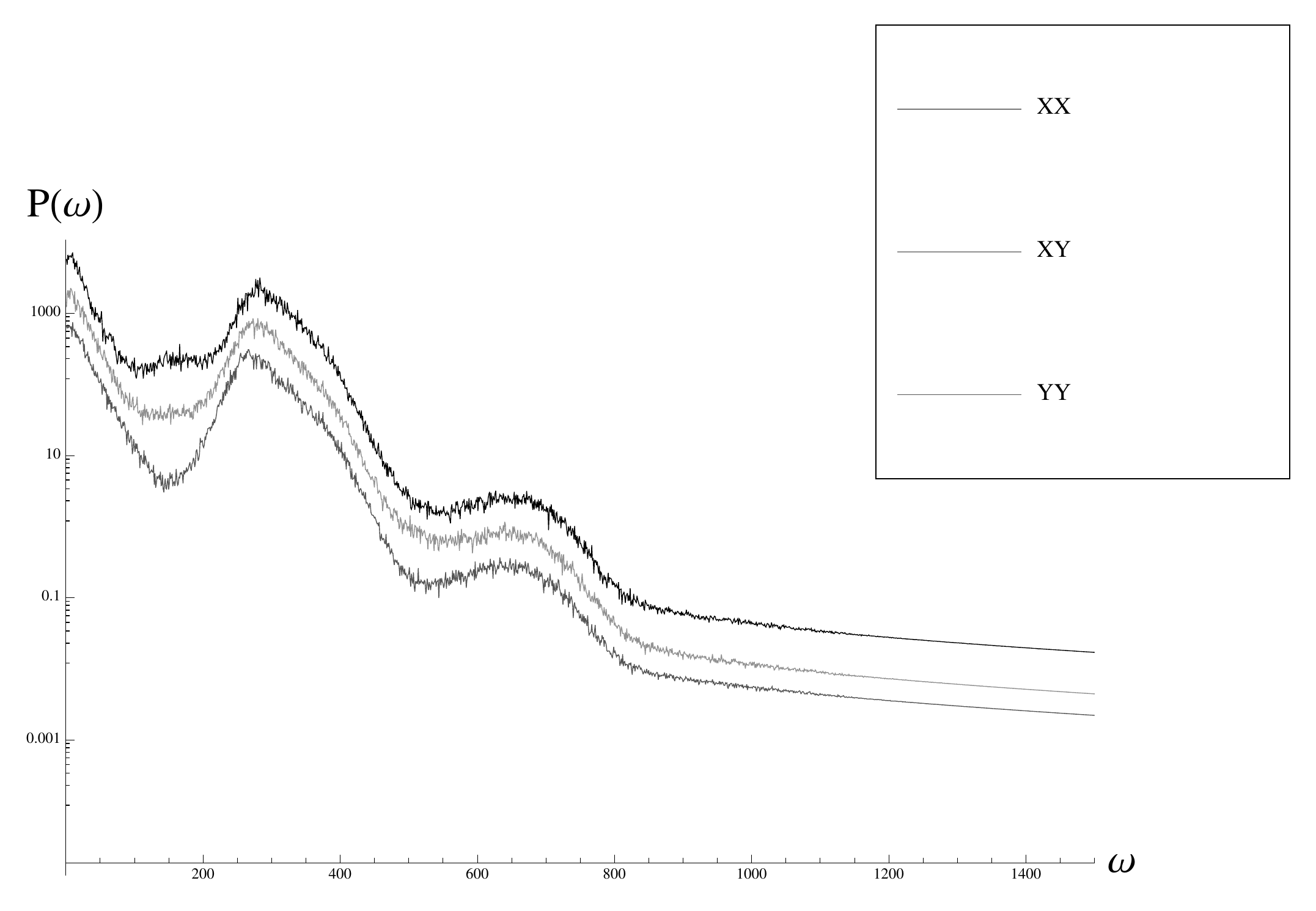}
\caption{Power spectra in arbitrary units for $\CO_2$ for various matrix combinations. The plots are artificially separated so they can be easily distinguished. In most of these plots the net normalization of $\tr(X^2)\simeq \tr(Y^2)$ is very close to each other, as shown previously. These arise from $40\times 40$ matrices with initial velocity in our  initial conditions set to $v=100$.}
\label{fig:powbmn}
\end{figure}

We can also look at what happens when we deform the system from the BFSS matrix model to the BMN matrix model. 
It is interesting to see how symmetry breaking is implemented in the power spectra. This is depicted in Fig. \ref{fig:powbmn}.

As shown in the figure, the power spectrum of $\tr(X^1+iX^2)^2$ acquires a new bump near where the power spectra of the BFSS matrix model had a local minimum. The bump is less pronounced in the $\tr(XY)$ channel and seems absent in the $\tr(Y^1+iY^2)^2$ channel. There is also a deformation of the $\tr(XY)$ bump near the second minimum. This shows that three objects that had the same symmetry properties in the BFSS matrix model 
exhibit the broken symmetry of the BMN matrix model in their dynamics. If we zoom in near the bump at zero, we can also see small differences. 
The size of these bumps depends on the strength of the mass term in the BMN matrix model. 
A full analysis of such deformations would consider the mixing between modes in different symmetry classes and would provide some understanding of response theory beyond linear response: we made a finite deformation of the lagrangian and the response to the deformation 
can be measured in dynamical quantities.

Understanding how the bump size depends on $N$ and the effective mass, keeping the temperature fixed, can give us a better understanding of the phase diagram of the BMN matrix model and can let us fine tune the system to obtain an appropriate large $N$ limit. 
This requires the effects of the mass term
and cubic term deformations to be compatible with the large $N$ scaling we obtained in Sec. \ref{sec:thermal}.
We can analyze this in terms of their expected contributions to the free energy. 
We expect the corrections to the free energy from the mass terms to be of order $N \mu^2 N^{1/2} T^{1/2}$,
as compared with $N^2 T$. If we want the ratios of these two contributions to the free energy to stay fixed (so that we get a proper large $N$ counting of the free energy), we need to scale $
\mu^2\simeq N^{1/2} T_0^{1/2}$ for some reference temperature $T_0$.  Performing such an analysis is beyond the scope of the present paper.

\section{Factorization}
\label{sec:fac}

A crucial aspect of large $N$ physics is factorization.
This states that correlators have a large $N$ expansion in powers of $1/N$, where the leading power of $N$ arises from planar diagrams \cite{'tHooft:1973jz} and subleading corrections arise from higher genus Feynman diagrams.
For the simplest observables, the leading expectation value of a product of observables is the product of the expectation values, so long as these expectation values do not vanish in the first place.
The arguments of planarity, or more precisely, that many of the simplest excitations in such systems give rise to an approximately free theory, with interactions governed by $1/N$ or $1/\sqrt{N}$ corrections 
 is an integral part of gravitational holography \cite{Maldacena:1997re} (see \cite{Witten:1979kh} for a nice description of this physics).

The classical dynamics we have factorizes in a trivial sense: the value of  a product of any set of observables at time $t$ is the product of the values.
However, what we would like to check is that expectation values that are averaged over time have this property as well:
that the simple degrees of freedom can be converted into approximately `free' constituents.
The simplest way to think about this is that the thermal fluctuations in observables are independent of each other for the factorized degrees of freedom, and because the degrees of freedom are approximately free, the fluctuations should be gaussian.
In the quantum theory near the vacuum, there is a standard way to understand that this leads to a consistent large $N$ classical dynamics \cite{Yaffe:1981vf}.
Here we want to check that there are also classical thermodynamic (or more precisely hydrodynamic) variables on which one can do a similar type of analysis. 

Imagine that the simulations we are doing with time evolution in the BFSS or BMN matrix model can be reinterpreted as a matrix model calculation 
\begin{equation}
\langle \CO(t) \rangle_t \simeq  \frac{\int \CO(X) \exp(-\beta V(X))_{\text{MM}}}{\int\exp(-\beta V(X))_{\text{MM}}}
\label{eq:hdmc}
\end{equation}
If the right hand side factorizes, then so does the left hand side.
Indeed, the algorithm we are following would correspond to a hydrodynamic Monte Carlo code to compute the right hand side.
So long as the trajectories in the system we have are sufficiently mixing, then the left hand side and the right hand side should match for a long enough $t$. 

What should be interesting to notice is that the right hand side of Eq. \eqref{eq:hdmc} in the BFSS matrix model has no quadratic term.
Hence the usual argumentation based on planar diagrams does not hold, as there are no quadratic terms in the $X$ variables.
One can also argue that in the BMN matrix model at large $\beta$ the quadratic terms matter very little, and it is instead the quartic term that dominates.
Moreover, the BFSS potential also has flat directions.
Both of these observations combined could conceivably produce anomalous powers of $N$ in the final answer, so it is worth checking that factorization holds.

\begin{figure}[ht]
\includegraphics[width=3.5 in]{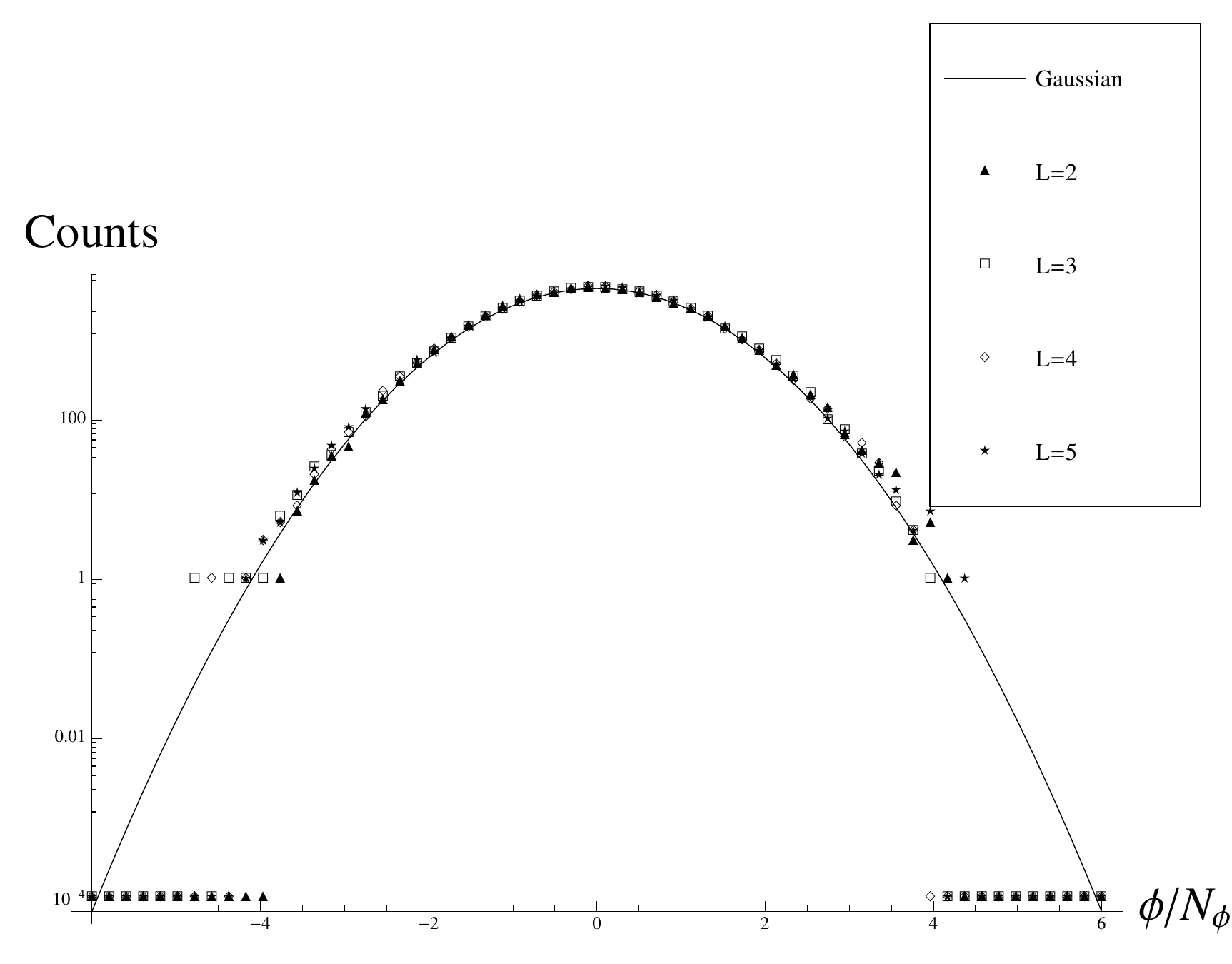}
\caption{Test of factorization for $N=87$.
We bin the samples of $\Real(\CO_L(t))$ for various $L$ obtained from the time series after thermalization.
We compare to a Gaussian model of the data.
Using logarithmic scaling in the counts permits us to check the tails of the distribution.
We put measured counts of zero at $10^{-4}$.}
\label{fig:gaussian}
\end{figure}

We will proceed in two steps.
First, we will check some consequences of factorization at some large value of $N$.
For example, consider the matrix model correlators (as in Eq. \eqref{eq:hdmc}) of the following form
\begin{equation}
\langle \CO_L^n \bar \CO_L^m \rangle = A^L_{m,n} 
\end{equation} 
where $\CO_L = \tr(Z^L)$, and $\bar \CO_L = \tr(\bar Z^L)$, for the case $L \geq 2$ and we take $Z = X^1 + iX^2$, or any of its rotations.
Rotational invariance of the ensemble implies that $A_{m,n} = A_m \delta_{m,n}$.
Our goal is to understand the $A^L_m$ at large $N$.
Notice that $\langle \CO_L\rangle = 0$, so this is exactly one of the cases where the naive large $N$ factorization does not apply.
Instead, we can consider $A^L_1$ as our first non-trivial value, and use it to normalize the answers.
We expect that because the potential in the BFSS matrix model is a scaling function, that the ratios $A^L_m/A^L_1$ are independent of the effective coupling constant $\beta$. 
Arguing analogously to \cite{BMN}, we can think of $\tr(Z^L)$ as a raising operator for a composite field (an {\em in} single string state), and $\tr(\bar Z^L)$ as the corresponding lowering operator (an {\em out} single string state).
The effective propagator for raising and lowering would be just $A_1$.
This is the naive argumentation if planar diagrams were applicable.
Then we would find that 
\begin{equation}
A_m = m! (A_1)^m
\end{equation}
from all the free contractions between raising and lowering operators.
This would be the leading diagram for closed string propagation without interactions, and furthermore, other diagrams with interactions would be suppressed by $1/N^2$. 
Thus the statistical distribution of $\tr(Z^L)$ would be that of a random Gaussian variable.
A simple check is to see if this is correct is to bin the results of sampling the real part of $\tr(Z^L)$, divided by its normalization (which in this case is $A^L_1/\sqrt 2$) and to compare it to a Gaussian model for the distribution normalized to the number of samples. 
This is independent of $L$.
We can see the results of this procedure in Fig. \ref{fig:gaussian}.

The results of the test are that the different $\CO_L$ have Gaussian statistics 
and we conlcude that the correlators do factorize in this sense.
We did this for $N=87$, but the results are very similar for other values of $N$.

We also consider correlators such as 
\begin{equation}
\frac{\langle \CO_L \CO_M \bar\CO_{L+M}\rangle} { \sqrt{A_1^L A_1^M A_1^{L+M}}}\sim \frac{ C_{L, M, L+M}}{N}+ O(1/N^3),
\label{eq:Clm}
\end{equation}
which should give rise to structure constants $C$ that have a well defined large $N$ limit.
If we ignore the $1/N^3$ corrections, then the $C$ should be independent of $N$ up to statistical uncertainties.
Rotation invariance implies that the correlators above are real, since $\bar Z$ can be obtained by an $\text{SO}(9)$ rotation of $Z$.

Note that this correlator decays only $1/N$ and not as $1/N^2$.
This is important for understanding the large error bars of the measurement \footnote{Similar issues appeared in \cite{Berenstein:2008jn}, where an object similar to $C_{L,M,L+M}$ had a theory prediction that was being tested against a model.}. The value of an instantaneous measurement on the left is of order one while the expectation value is of order $1/N$, thus the various measurements must cancel each other most of the time, leaving a small residual. 
A non-zero average could be also called a ``violation of Gaussianity" if we think of the $\CO_L$ as statistically independent variables.
\begin{table}
\centering
\begin{tabular}{|c|c|c|c|c|c|c|c|c|}
\hline
& $N=10$  & $N=13$ & $N=18$ &$N=87$  \\
\hline
$C_{2,2,4}$& $4.97\pm 0.51$ & $4.54\pm 0.15$ & $4.94\pm 0.8$ &$4.97\pm 1.2$\\
$C_{3,3,6}$ & $6.97\pm 0.86$ & $6.9\pm 0.4 $&$7.58\pm 0.5$&$8.36\pm1.4$\\
\hline
\end{tabular}
\caption{Values of $C_{L,M, L+M}$ at various values of $N$}
\label{tab:clm}
\end{table}
A simple test for two possible $C$ is shown in table \ref{tab:clm}, where we can see that the $C$ are indeed $N$ independent given the error bars.
This gives us confidence that the standard large $N$ counting is applicable.

\begin{figure}[ht]
\includegraphics[width=3.5 in]{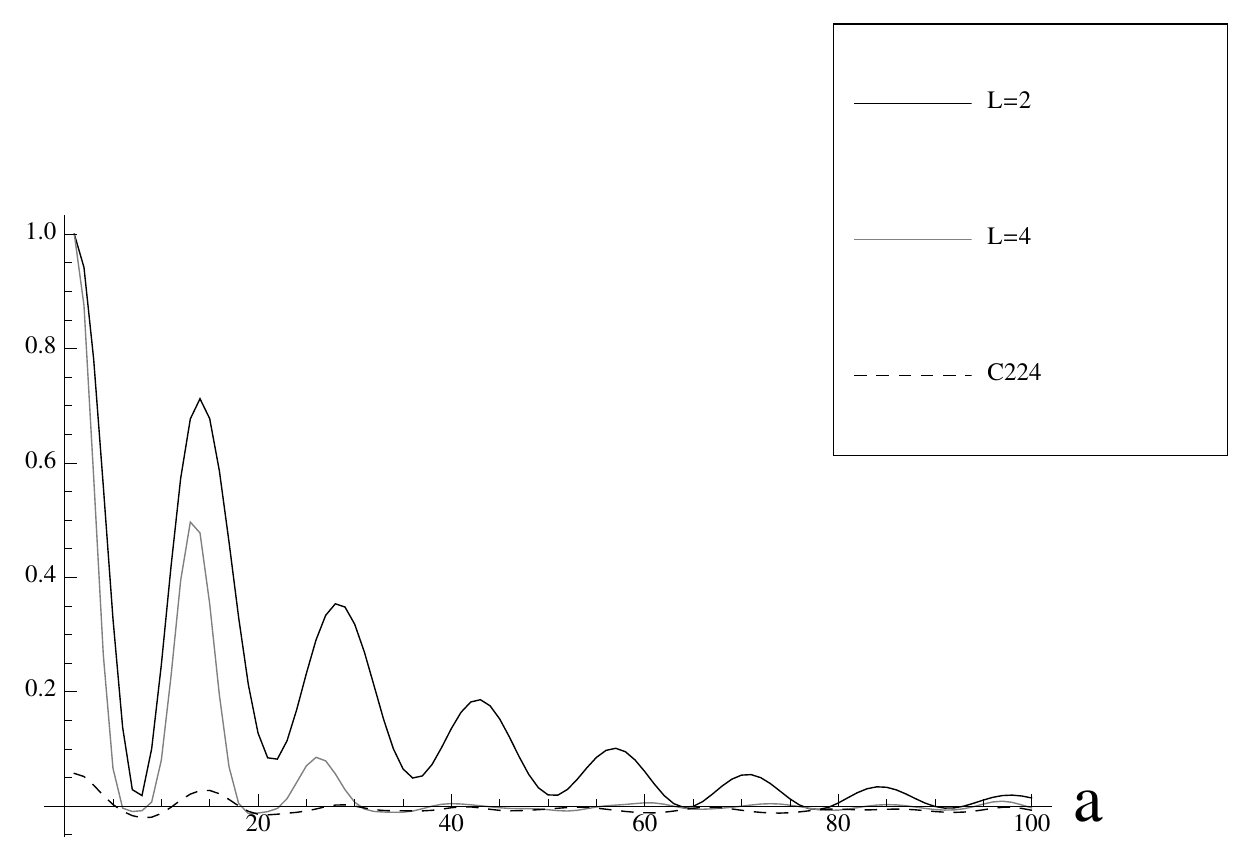}
\caption{Correlation function $C_{2,2,4}(a)/N$ compared to the normalized autocorrelations $A^{2,4}_1(a) = \langle \tr(Z^{2,4}(t)) \tr(\bar Z^{2,4}(t+a))\rangle_t $.
The normalization is $A_1^{2,4}=A_1^{2,4}(0)$ for $L=2,4$ respectively.
The statistical error band on $C_{2,2,4}$ should be roughly at $25\%$ of the maximum around the value at zero.
The graph shows data from a run at $N=87$.}
\label{fig:correlation}
\end{figure}

We can generalize Eq. \eqref{eq:Clm} to include time dependence 
and check that
\begin{equation}
\label{eq:Clmt}
\frac{\langle \CO_L(t) \CO_M(t) \bar\CO_{L+M}(t+a)\rangle_t} { \sqrt{A_1^L A_1^M A_1^{L+M}}}\sim 
\frac{ C_{L, M, L+M}(a)}{N}+ O(1/N^3),
\end{equation}
where now the $C_{L,M, L+M}$ indicate nonlinear correlations with time dependence.
If the matrix model and gravity are to be matched, these powers of $N$ should be robust.
We expect that the right hand side will decay with time $a$, as correlations typically do in chaotic systems.
A plot of the correlation function $C_{2, 2, 4}(a) N^{-1}$ can be seen in Fig. \ref{fig:correlation}, where it decays as expected.
The statistical error band that one should associate to the graph is similar in size to the error bars seen in Table \ref{tab:clm}. 

Note how the correlation between the different variables $\CO_L(t)$ peaks when the autocorrelation of the variables also peaks and that the correlations between different variables roughly decay as the autocorrelation function decays.
This is consistent with naive expectations.
The purpose of this check is to show that large $N$ counting is also applicable to general dynamical questions: if the initial non-gaussianity is of order $1/N$ and it bounds the time dependent non-gaussianity, then these can not be larger than $1/N$.

\section{Gravitational interpretation}
\label{sec:grav-int}

In this section we discuss the extent to which we can call the thermalized classical dynamics of the matrix model a black hole.
Let us begin with the D-brane background geometries in the absence of excitations.
They are characterized by the supergravity solutions found in \cite{Horowitz:1991cd}.
In the string frame the ten dimensional metric is given by
\begin{equation}
ds^2 = H^{-1/2}(r) (dx_{||}^2) + H^{1/2}(r)(dr^2 + r^2 d\Omega_{8-p}^2),
\end{equation}
where $dx_{||}$ are the $p+1$ coordinates that run along the worldvolume of the $p$-brane, $r$ is a radial direction and $\Omega_{8-p}$ is an $8-p$ dimensional sphere, and $H$ is a harmonic form given by
\begin{equation}
\label{eq:Hform}
H(r) = 1 + \frac{Na}{r^{7 - p}}
\end{equation}
where $a$ is a constant that depends on $p$, but not on $N$ or $r$.

We start with a few comments on string theory in this geometry. 
For $p\leq 3$ these metrics produce a long throat near $r=0$, in that $\int_0^R H^{1/4}(r)\, dr = \infty$.
If not for the time warping, a string that stretches to the brane, at the origin, would have infinite mass.
This means the region near $r=0$ can be considered a large volume, in string units, 
and we are justified in dropping the $1$ in \eqref{eq:Hform}.
As noted in \cite{Itzhaki:1998dd}, the string coupling remains finite as we take the near horizon, 
decoupling limit. 
For $p=0$, which is the case of interest given the $\text{SO}(9)$ symmetry of the BFSS system, 
the effective curvature becomes large for large $r$, and is small near $r=0$.

Now imagine adding energy to this system and creating a black hole with the same asymptotics as this background. 
These black holes have positive specific heat. 
If the temperature is low, the curvature of the black hole near the horizon is small in string units \cite{Itzhaki:1998dd}. Because the dilaton runs in these geometries towards small coupling in the UV, 
the effective string scale depends on the position, and the curvature becomes large in string units in the UV. 
Correspondingly, the curvature will be  large in string units if the temperature is high, so the high temperature black hole is stringy.
The regime of interest for us is high temperature, where the curvature near the putative back hole is large and where stringy corrections must be important. 
In the spirit of \cite{Horowitz:1996nw}, we should be able to describe that region by replacing the black holes by configurations of D-branes, because we are in the stringy regime. The classical dynamics we studied is the 
microscopic, classical dynamics of the D-brane configurations.  
These configurations with D-brane sources are  horizonless in the sense of classical gravity,
but when the system cools enough we recover a black hole. This notion that we recover a black hole when we cool the system down
 is the reverse of the scenario in \cite{Horowitz:1996nw}, where the authors argued that a black hole that becomes small enough to become stringy is replaced by a set of strings and possibly D-branes. We assume that this philosophy is applicable in this case as well, as there is no reason to expect a phase transition and we can corroborate this by examining the scalings of various quantities with respect to $N$ and $T$.



From our calculations via the virial theorem, we know that the radius of the brane configurations in the classical system grows as $R\simeq N^{1/4} T^{1/4}$, so the effective (codimension two) size covered by the brane system scales as $R^8 \simeq N^2 T^2$. The energy is of order $N^2 T$ whereas the entropy of the D-brane gas is of order $N^2 \log (T/T_0)$, where $T_0$ is some reference temperature whose precise value doesn't concern us here.
As computed in \cite{Catterall:2007fp}, for low temperatures the entropy scales as $S\simeq N^2 T^{9/5}$ and the free energy scales as $E\simeq N^2 T^{14/5}$.
The smooth transition to the brane gas occurs near $T\simeq 1$ in their units.
Since for the black brane we have the usual area law for the entropy $S \simeq r^8$, for $T \simeq 1$ we have
 $R^8 \simeq r^8$. So the brane gas extends all the way to where we would imagine the black hole horizon to be. 
The entropy in both cases is of order $N^2$, as well as their energy, so there is no gap between the energy or entropy scalings that would suggest a phase transition. This is the essence of the argument in \cite{Horowitz:1996nw} for smooth interpolation between strings and black hole physics.
Furthermore, that the configuration reaches the same radius as where the black hole horizon would be located is very similar to the fuzzball geometries (see \cite{Mathur:2005zp} for a review).
This suggests that the system can be described both as a black hole and as a collection of D-branes.
If the transition between the descriptions is smooth, we can choose one or the other depending on the question we want to ask.

Now we want to consider the dissipation we observe, in the sense of decaying correlations, in our simulations.
As shown in our Fig. \ref{fig:collapse}, for all $N$ we get a very similar power spectrum of fluctuations.
Applying fluctuation-dissipation relations, we can use these to characterize response functions.
If we were to match to gravity, valid at lower temperatures,
we would expect dissipation to be related to the presence of quasinormal modes 
associated with a black hole horizon. 

We expect the point of comparison between both descriptions to be the analytic structure of correlation functions and power spectra, such as those shown in Fig. \ref{fig:collapse}.
As shown in the figure, there are many suggestive straight lines that describe the logarithm of the power spectrum.
Remember also that the power spectrum is symmetric about $\omega =0$.
The most naive fit to the graph near $\omega \simeq 0$ is 
\begin{equation}
P(\omega) \simeq \exp( -\beta |\omega|).
\label{eq:pwspec1}
\end{equation}
The absolute value is not an analytic function of the complex variable $\omega$ and the Fourier transform of Eq. \eqref{eq:pwspec1} only decays polynomially at large times.
If the singularity is smoothed out very near $\omega = 0$, one can imagine that an analytic function of $\omega$ replaces $|\omega|$.
The simplest such function is $f(\omega) =\sqrt {\omega^2+\epsilon}$. When $\epsilon \to 0$ we recover the result above.
The presence of $\epsilon$ suggests a pair of branch cuts beginning near $\omega\simeq 0$ along the imaginary axis.
If these are the closest singularities to the real axis, then the function decays exponentially in time.
Thus, even if the ultimate late time fate of the system has correlation functions that  decay exponentially in time, there can be a long transient where the decays of the correlation functions are only polynomial.

A square root branch cut can be approximated by a density of poles (this is often seen in matrix models \cite{Brezin:1977sv}).
Since we only have the analytic function on the real numbers, extrapolating to find the poles requires a very good understanding of the analytic structure of the function and the pattern of pole locations.
Just for comparison, another  similar function would be given by
\begin{equation}
P(\omega) \simeq \exp\left( -\sqrt{ \beta \omega^2 + \epsilon} -\sqrt{ \beta^* \omega^2 + \epsilon^*}\right),
\label{eq:pwspec2}
\end{equation}
where $\beta$, $\beta^*$ are complex conjugates of each other.
This is real on the real axis and has four branch cuts near the origin starting at $\omega= \pm i \sqrt{\epsilon/\beta}$ and $\omega= \pm i \sqrt{\epsilon^*/\beta^*}$.
Such branch cuts could indicate a series of poles along straight lines, starting from the brach cut endpoints and going towards infinity.
Such patterns of quasinormal modes aligning on fairly straight curves have been observed in Schwarzschild and AdS black holes 
\cite{Kokkotas:1999bd, Berti:2009kk, Konoplya:2011qq}.
A particularly nice set of  examples can  be found in \cite{Kovtun:2005ev}, Fig. 4, and \cite{Marolf:2012dr}. In gravity setups such patterns are interpreted in terms of the membrane paradigm \cite{Iqbal:2008by}.

Our results suggest that there is agreement between our analytics and the existence of a large collection of quasinormal modes for each $L$ near $\omega= 0$.
We might eventually be able to interpret them as shear or sound modes once we understand the details of how these modes are mapped into each other. 
Notice that as we increase $L$, as in Fig. \ref{fig:pow27}, the curves become flatter near $\omega= 0$ for even $L$, suggesting that the corresponding poles nearest to the origin are moving away from the real axis.
This suggests that there is a dispersion relation for the frequencies $\omega(L)$ of 
these poles such that $\Imag(\omega(L))$ increases with $L$.
Here $L$ is angular momentum about the sphere of the spherical black hole geometry, so such a dispersion relation could be interpreted geometrically, but we haven't made this precise yet: in the absence of a theory of where these poles 
and branch cuts should be located, any  match to a dispersion relation would be mere speculation. 

The analytic structure of the power spectra of observables in a chaotic dynamical system is generally 
controlled by poles off the real axis known as Pollicot-Ruelle resonances. The locations of these are known for some 
simple systems, but finding them in general, even for very low dimensional systems, is difficult 
(see, e.g., \cite{Florido}). For large systems with many degrees of freedom there are expected to be 
accumulations of poles and also
branch cuts \cite{Gaspard-book}, in line with our discussion above. 
More detailed information on the analytic structure may be difficult to obtain, 
but we plan to analyze this in more detail in the future. 
We also note the other lines drawn in the Fig. \ref{fig:collapse}, which suggest similar interpretations for the other peaks as modes 
whose frequencies begin with a non-zero real part. 

We do not have a theory for the analytic structure of the power spectra, but
based on the 
data and our general understanding of dissipative phenomena
we have speculated that there are branch cuts in the analytically continued spectra.
Given the behavior of black hole horizons, as in the membrane paradigm and as characterized 
by the quasinormal modes,
we think our numerical data are consistent with having a smooth transition 
from the matrix configurations to a black hole at low temperatures.

\section{Conclusion}
\label{sec:conclusion}

Let us first summarize our observations from our numerical simulations of the BMN and BFSS matrix models.
The dynamics of the matrix models are chaotic. 
There are strong indications that the systems thermalize, most prominently that the typical matrices 
of the momentum and position variables behave, at late times, as random matrices from the traceless GUE ensemble. 
This is expected for the momentum matrices because the Hamiltonian is quadratic in the relevant degrees of freedom, 
but because of the nonlinearities in the potential we would not immediately guess this for the position matrices. 
We also showed how the presence of constraints alters the naive arguments about the 
appropriate random matrix ensembles for these systems.
We have seen that certain observables behave hydrodynamically, i.e., 
their power spectra at equilibrium are approximately independent of the total number of degrees of freedom 
and they have approximately gaussian statistics.
We have also seen that large $N$ counting applies to time correlation functions between observables, so that violations of gaussianity scale like $1/N$, i.e., 
there is factorization of these degrees of freedom. 
This $1/N$ scaling is associated with quantum corrections under the usual AdS/CFT power counting arguments.

These particular matrix models are interesting because of their dual gravitational interpretation. 
Though quantum effects are crucial to the emergent gravitational dynamics in general, we have argued that 
at high temperatures the classical dynamics of these matrix models do encode certain dual gravitational phenomena, 
including black holes, albeit in a stringy regime. As such, a direct comparison with supergravity 
solutions is not possible, but we can look for and have found some qualitative agreement. 

The first level of correspondence is between black holes, broadly defined to include large stringy  corrections, and equilibrium thermal states in these models. 
This thermodynamic correspondence 
was referred to in both of the seminal papers on BFSS matrix theory and AdS/CFT \cite{BFSS, Maldacena:1997re}
and subsequently investigated and verified in many situations.
The main work of this paper has been the identification of high temperature equilibrium configurations 
in the matrix models and some investigation of their equations of state and fluctuations, 
in the manner of a molecular dynamics simulation. This is a continuation of the work 
first presented in \cite{Asplund:2011qj}.
As a first step toward a more detailed correspondence,
we have presented evidence here that the approximate analytic structure we find in the power spectra 
of certain observables at large $N$
could be the remnant of the quasinormal modes of supergravity black holes. 
This is suggested by the numerical data, where the results we find on the real axis seem to be well approximated by
functions that, when analytically continued, would have branch cuts in the complex plane. 
These could be an approximation of a sequence of roughly evenly spaced poles.  

It is possible to study the geometry that emerges from these matrix models in more detail, 
for example by using the methods presented recently in \cite{Berenstein:2012ts}. 
Such geometric data is based on probe branes and the fermionic degrees of freedom that connect them 
to the configurations we have shown here. This information would complement what is done here very nicely 
and we can speculate that 
in the future such reasoning will lead us to understand the emergence of gravitational horizons from such dynamics better.

We also seek to investigate the dissipation and other transport properties further. 
For example, the time autocorrelation functions of various observables, such as those displayed in 
Fig. \ref{fig:correlation}, can be integrated to compute the associated transport coefficients, via 
the Green-Kubo relations. We can then investigate how these depend on the temperature, 
$N$, or other variables, and see whether the behavior can be interpreted holographically 
and meets our expectations from gravity. 
We note that these transport phenomena are related to the chaotic dynamics of the system, 
characterized by quantities
such as the Lyapunov exponents, Pollicot-Ruelle resonances, and Kolmogorov-Sinai entropy
(see, e.g., \cite{2006PhyA..369..201G} for a review).
These quantities also control the far from equilibrium dynamics. In that vein, 
we may be able to profitably study 
other non-equilibrium phenomena
by applying further techniques and results from non-equilibrium statistical mechanics.

Reflecting on what we have accomplished here, we have made a modest attempt at reconciling 
gravitational dynamics in holographic setups with the theory of chaotic dynamical systems. 
We have argued that these types of analyses do cover interesting black holes in the stringy regime, 
which would mean the study of these situations is forced on us if 
we want to be complete.  From this vantage point, we have barely scratched the surface of what can be computed and analyzed and it would be interesting to pursue this further to improve our understanding of holography. 
These simulations of the classical dynamics provide new ways to address questions about black holes in simple systems where the numerical computations 
are easily implemented. It is also important to understand 
how to move away from the correspondence limit, which would mean quantum dynamics starts becoming important and the dynamics of fermions starts affecting the results. In such setups the tools of quantum chaos will play an increasingly important role. 
This is closely related to the question of whether at low temperatures it is the fermionic degrees of freedom that dominate, or if they only play a marginal role in most of the dynamical regimes of interest.

\begin{acknowledgments}

We would like to thank J. Hartle, G. Horowitz, C. Maes, 
D. Marolf, J.  Polchisnki, J. Santos, M. Srednicki, and E. Shuryak
for various discussions related to this work.
We thank J. Mangual for his discussions about random matrices 
and for his pursuit of a formula for the level density for the TGUE at arbitrary $N$. 
This led to a discussion on the website mathoverflow.net and the answer 
of F. Bornemann \cite{MO87019}, which ultimately 
resulted in the formula reported in \cite{Ho-Kahn:2011}. 
Work supported in part by DOE under grant DE-FG02-91ER40618 and by the Department of Energy Office of Science Graduate Fellowship Program (DOE SCGF), made possible in part by the American Recovery and Reinvestment Act of 2009, administered by ORISE-ORAU under contract no. DE-AC05-06OR23100.
CA has been supported in part by the FWO - Vlaanderen, Project No.
G.0651.11 and the Odysseus program, by the Federal Office for Scientific, Technical and
Cultural Affairs through the ``Interuniversity Attraction Poles Programme Belgian Science Policy" P7/37,
the European Science Foundation Holograv Network and by a grant from the John Templeton Foundation. 
The opinions expressed in this publication are those of the authors and do not necessarily reflect the views of the 
John Templeton Foundation.
\end{acknowledgments}

\appendix

\section{Measuring the Temperature}
\label{sec:temp}

Here we derive the relation between the temperature of a system with constraints and the second moments of its degrees of freedom.
Consider a system with $2D$ degrees of freedom $\vec{x}$, $\vec{p}$ with the standard kinetic energy, potential $V(\vec{x})$, and $k$ constraints $C^i(\vec{x}, \vec{p}) = 0$ with each $C^i$ linear in the momenta.
The canonical partition function is given by
\newcommand{\Zcal}{ {\mathcal{Z}} }
\begin{equation}
\Zcal = \int d^Dx d^Dp \, \prod_{i=1}^k\delta(C_i(\vec{x}, \vec{p})) \exp\left[-\beta\left( \frac{1}{2}|\vec{p}|^2 + V(\vec{x})\right)\right]
\end{equation}
We scale the momenta by a parameter $\sqrt{\gamma}$ and then rescale the $\delta$ functions by the inverse of that parameter.
Since the constraints are linear in the momenta
\begin{equation}
\Zcal = \gamma^{(D - k) / 2}\int d^Dx d^Dp \, \prod_{i=1}^k\delta(C_i(\vec{x}, \vec{p})) \exp\left[-\beta\left( \frac{\gamma}{2}|\vec{p}|^2 + V(\vec{x})\right)\right]
\end{equation}
The partition function is independent of $\gamma$ since all we have done is rescaled the momenta.
Differentiating we have
\begin{align}
0 = \frac{\partial\Zcal}{\partial\gamma} &= \frac{D - k}{2\gamma}\gamma^{(D - k) / 2}\int d^Dx d^Dp \, \prod_{i=1}^k\delta(C_i(\vec{x}, \vec{p})) \exp\left[-\beta\left( \frac{\gamma}{2}|\vec{p}|^2 + V(\vec{x})\right)\right] \nonumber \\
&\qquad - \frac{\beta}{2}\gamma^{(D - k) / 2}\int d^Dx d^Dp \, |\vec{p}|^2 \prod_{i=1}^k\delta(C_i(\vec{x}, \vec{p})) \exp\left[-\beta\left( \frac{\gamma}{2}|\vec{p}|^2 + V(\vec{x})\right)\right]
\end{align}
Letting $\gamma = 1$ we have
\begin{equation}
0 = \frac{D - k}{2}\Zcal - \frac{\beta}{2}\Zcal\langle |\vec{p}|^2\rangle \Rightarrow (D - k)T = \langle |\vec{p}|^2\rangle
\end{equation}
In order to measure the temperature properly, we must subtract off the number of constraints from the degrees of freedom.


\begin{thebibliography}{99}

\bibitem{Policastro:2001yc} 
  G.~Policastro, D.~T.~Son and A.~O.~Starinets,
  ``Shear viscosity of strongly coupled $\mathcal{N}=4$ supersymmetric Yang-Mills plasma,''
  Phys.\ Rev.\ Lett.\  {\bf 87}, 081601 (2001)
  [hep-th/0104066].
  
\bibitem{deForcrand:2010ys} 
  P.~de Forcrand,
  ``Simulating QCD at finite density,''
  PoS LAT {\bf 2009}, 010 (2009)
  [arXiv:1005.0539 [hep-lat]].
 
\bibitem{Kovtun:2004de} 
  P.~Kovtun, D.~T.~Son and A.~O.~Starinets,
  ``Viscosity in strongly interacting quantum field theories from black hole physics,''
  Phys.\ Rev.\ Lett.\  {\bf 94}, 111601 (2005)
  [hep-th/0405231].
  
\bibitem{MolecDyn}
D. Frenkel and B. Smit. \emph{Understanding Molecular Simulation. 2nd Ed.} Academic Press: San Diego. (2002) 

\bibitem{BFSS} 
  T.~Banks, W.~Fischler, S.~H.~Shenker and L.~Susskind,
  ``M theory as a matrix model: A Conjecture,''
  Phys.\ Rev.\ D {\bf 55}, 5112 (1997)
  [hep-th/9610043].
  
\bibitem{BMN} 
  D.~E.~Berenstein, J.~M.~Maldacena and H.~S.~Nastase,
  ``Strings in flat space and pp waves from $\mathcal{N} =4$ super Yang-Mills,''
  JHEP {\bf 0204}, 013 (2002)
  [hep-th/0202021].
 
\bibitem{Itzhaki:1998dd} 
  N.~Itzhaki, J.~M.~Maldacena, J.~Sonnenschein and S.~Yankielowicz,
  ``Supergravity and the large $N$ limit of theories with sixteen supercharges,''
  Phys.\ Rev.\ D {\bf 58}, 046004 (1998)
  [hep-th/9802042].
  
\bibitem{Berenstein:2002sa} 
  D.~Berenstein and H.~Nastase,
 ``On light cone string field theory from super Yang-Mills and holography,''
  hep-th/0205048.
  
\bibitem{Marolf:2002ye} 
  D.~Marolf and S.~F.~Ross,
  ``Plane waves: To infinity and beyond!,''
  Class.\ Quant.\ Grav.\  {\bf 19}, 6289 (2002)
  [hep-th/0208197].
  
  
\bibitem{Horowitz:1996nw} 
  G.~T.~Horowitz and J.~Polchinski,
  ``A Correspondence principle for black holes and strings,''
  Phys.\ Rev.\ D {\bf 55}, 6189 (1997)
  [hep-th/9612146].

\bibitem{Festuccia:2006sa} 
  G.~Festuccia and H.~Liu,
  ``The Arrow of time, black holes, and quantum mixing of large $N$ Yang-Mills theories,''
  JHEP {\bf 0712}, 027 (2007)
  [hep-th/0611098].

\bibitem{Sekino:2008he} 
  Y.~Sekino and L.~Susskind,
  ``Fast Scramblers,''
  JHEP {\bf 0810}, 065 (2008)
  [arXiv:0808.2096 [hep-th]].

\bibitem{Asplund:2011qj} 
  C.~Asplund, D.~Berenstein and D.~Trancanelli,
  ``Evidence for fast thermalization in the plane-wave matrix model,''
  Phys.\ Rev.\ Lett.\  {\bf 107}, 171602 (2011)
  [arXiv:1104.5469 [hep-th]].
  
\bibitem{Riggins:2012qt} 
  P.~Riggins and V.~Sahakian,
  ``On black hole thermalization, D0 brane dynamics, and emergent spacetime,''
  Phys.\ Rev.\ D {\bf 86}, 046005 (2012)
  [arXiv:1205.3847 [hep-th]].

\bibitem{Edalati:2012jj} 
  M.~Edalati, W.~Fischler, J.~F.~Pedraza and W.~Tangarife Garcia,
  ``Fast Scramblers and Non-commutative Gauge Theories,''
  JHEP {\bf 1207}, 043 (2012)
  [arXiv:1204.5748 [hep-th]].

  
\bibitem{Lashkari:2011yi} 
  N.~Lashkari, D.~Stanford, M.~Hastings, T.~Osborne and P.~Hayden,
  ``Towards the Fast Scrambling Conjecture,''
  arXiv:1111.6580 [hep-th].

\bibitem{Barbon:2011pn} 
  J.~L.~F.~Barbon and J.~M.~Magan,
  ``Chaotic Fast Scrambling At Black Holes,''
  Phys.\ Rev.\ D {\bf 84}, 106012 (2011)
  [arXiv:1105.2581 [hep-th]].

  
\bibitem{Barbon:2011nj} 
  J.~L.~F.~Barbon and J.~M.~Magan,
  ``Fast Scramblers Of Small Size,''
  JHEP {\bf 1110}, 035 (2011)
  [arXiv:1106.4786 [hep-th]].
  
\bibitem{Barbon:2012zv} 
  J.~L.~F.~Barbon and J.~M.~Magan,
  ``Fast Scramblers, Horizons and Expander Graphs,''
  JHEP {\bf 1208}, 016 (2012)
  [arXiv:1204.6435 [hep-th]].
 
\bibitem{Iizuka:2008hg} 
  N.~Iizuka and J.~Polchinski,
  ``A Matrix Model for Black Hole Thermalization,''
  JHEP {\bf 0810}, 028 (2008)
  [arXiv:0801.3657 [hep-th]].
  
\bibitem{Iizuka:2008eb} 
  N.~Iizuka, T.~Okuda and J.~Polchinski,
  ``Matrix Models for the Black Hole Information Paradox,''
  JHEP {\bf 1002}, 073 (2010)
  [arXiv:0808.0530 [hep-th]].

\bibitem{Maldacena:1997re} 
  J.~M.~Maldacena,
  ``The Large N limit of superconformal field theories and supergravity,''
  Adv.\ Theor.\ Math.\ Phys.\  {\bf 2}, 231 (1998)
  [hep-th/9711200].

\bibitem{Srednicki}
  M. ~ Srednicki, 
  ``Chaos and quantum thermalization"
  Phys.\ Rev.\ E {\bf 50},  888�1�71 (1994)
  [arXiv:cond-mat/9403051]



\bibitem{Dai:1989ua} 
  J.~Dai, R.~G.~Leigh and J.~Polchinski,
  ``New Connections Between String Theories,''
  Mod.\ Phys.\ Lett.\ A {\bf 4}, 2073 (1989).
  J.~Polchinski,
  ``Dirichlet Branes and Ramond-Ramond charges,''
  Phys.\ Rev.\ Lett.\  {\bf 75}, 4724 (1995)
  [hep-th/9510017].

\bibitem{Taylor:1998tv} 
  W.~Taylor and M.~Van Raamsdonk,
  ``Supergravity currents and linearized interactions for matrix theory configurations with fermionic backgrounds,''
  JHEP {\bf 9904}, 013 (1999)
  [hep-th/9812239].

\bibitem{DKPS} 
  M.~R.~Douglas, D.~N.~Kabat, P.~Pouliot and S.~H.~Shenker,
  ``D-branes and short distances in string theory,''
  Nucl.\ Phys.\ B {\bf 485}, 85 (1997)
  [hep-th/9608024].

\bibitem{BC} 
  D.~Berenstein and R.~Corrado,
  ``M(atrix) theory in various dimensions,''
  Phys.\ Lett.\ B {\bf 406}, 37 (1997)
  [hep-th/9702108].

\bibitem{Becker:1997xw} 
  K.~Becker, M.~Becker, J.~Polchinski and A.~A.~Tseytlin,
  ``Higher order graviton scattering in M(atrix) theory,''
  Phys.\ Rev.\ D {\bf 56}, 3174 (1997)
  [hep-th/9706072].

\bibitem{Verlinde:2010hp} 
  E.~P.~Verlinde,
  ``On the Origin of Gravity and the Laws of Newton,''
  JHEP {\bf 1104}, 029 (2011)
  [arXiv:1001.0785 [hep-th]].

\bibitem{Ho-Kahn:2011}
  K.~Ho, J.M.~Kahn,
  ``Statistics of Group Delays in Multimode Fiber with Strong Mode Coupling, Supplement,''
  Journal of Lightwave Technology, vol. 29, pp. 3119-3128, 2011
  [arXiv:1104.4527v2 [physics.optics]].
  See Supplement.

\bibitem{Catterall:2008yz} 
  S.~Catterall and T.~Wiseman,
  ``Black hole thermodynamics from simulations of lattice Yang-Mills theory,''
  Phys.\ Rev.\ D {\bf 78}, 041502 (2008)
  [arXiv:0803.4273 [hep-th]].

\bibitem{Hanada:2008ez} 
  M.~Hanada, Y.~Hyakutake, J.~Nishimura and S.~Takeuchi,
  ``Higher derivative corrections to black hole thermodynamics from supersymmetric matrix quantum mechanics,''
  Phys.\ Rev.\ Lett.\  {\bf 102}, 191602 (2009)
  [arXiv:0811.3102 [hep-th]].

\bibitem{Sav}
  G. Z.~ Baseyan, S. G.~ Matinyan and G. K. ~ Savvidi, JETP Lett. {\bf 29}, 585 (1979)

\bibitem{Chirikov:1981cm} 
  B.~V.~Chirikov and D.~L.~Shepelyansky,
  ``Stochastic Oscillation Of Classical Yang-mills Fields."
  JETP Lett.\  {\bf 34}, 163 (1981)
  [Pisma Zh.\ Eksp.\ Teor.\ Fiz.\  {\bf 34}, 171 (1981)].

\bibitem{Aref'eva:1997es} 
  I.~Y.~.Aref'eva, P.~B.~Medvedev, O.~A.~Rytchkov and I.~V.~Volovich,
  ``Chaos in M(atrix) theory,''
  Chaos Solitons Fractals {\bf 10}, 213 (1999)
  [hep-th/9710032].

\bibitem{Matinyan:1981ys} 
  S.~G.~Matinyan, G.~K.~Savvidy and N.~G.~Ter-Arutunian Savvidy,
  ``Stochasticity Of Classical Yang-mills Mechanics And Its Elimination By Higgs Mechanism. (in Russian),''
  JETP Lett.\  {\bf 34}, 590 (1981)
  [Pisma Zh.\ Eksp.\ Teor.\ Fiz.\  {\bf 34}, 613 (1981)].

\bibitem{Corley:2001zk} 
  S.~Corley, A.~Jevicki and S.~Ramgoolam,
  ``Exact correlators of giant gravitons from dual $\mathcal{N}=4$ SYM theory,''
  Adv.\ Theor.\ Math.\ Phys.\  {\bf 5}, 809 (2002)
  [hep-th/0111222].

\bibitem{Berenstein:2004kk} 
  D.~Berenstein,
  ``A Toy model for the AdS/CFT correspondence,''
  JHEP {\bf 0407}, 018 (2004)
  [hep-th/0403110].

\bibitem{Lin:2004nb} 
  H.~Lin, O.~Lunin and J.~M.~Maldacena,
  ``Bubbling AdS space and 1/2 BPS geometries,''
  JHEP {\bf 0410}, 025 (2004)
  [hep-th/0409174].
  
\bibitem{Das:1996wn} 
  S.~R.~Das and S.~D.~Mathur,
  ``Comparing decay rates for black holes and D-branes,''
  Nucl.\ Phys.\ B {\bf 478}, 561 (1996)
  [hep-th/9606185].

\bibitem{Klebanov:1997kc} 
  I.~R.~Klebanov,
  ``World volume approach to absorption by nondilatonic branes,''
  Nucl.\ Phys.\ B {\bf 496}, 231 (1997)
  [hep-th/9702076].

\bibitem{MWH}
  C. ~Manderfeld, J. ~ Weber and F.~ Haake,
  ``Classical versus quantum time evolution of (quasi) probability densities at limited phase-space resolution"
  J. Phys. A: Math. Gen. {\bf 34} (2001) 9893
  [arXiv:nlin/0107020]


\bibitem{Eckmann:1985zz} 
  J.~-P.~Eckmann and D.~Ruelle,
  ``Ergodic theory of chaos and strange attractors,''
  Rev.\ Mod.\ Phys.\  {\bf 57}, 617 (1985)
  [Addendum-ibid.\  {\bf 57}, 1115 (1985)].


\bibitem{'tHooft:1973jz} 
  G.~'t Hooft,
  ``A Planar Diagram Theory for Strong Interactions,''
  Nucl.\ Phys.\ B {\bf 72}, 461 (1974).

\bibitem{Witten:1979kh} 
  E.~Witten,
  ``Baryons in the 1/n Expansion,''
  Nucl.\ Phys.\ B {\bf 160}, 57 (1979).

\bibitem{Yaffe:1981vf} 
  L.~G.~Yaffe,
  ``Large n Limits as Classical Mechanics,''
  Rev.\ Mod.\ Phys.\  {\bf 54}, 407 (1982).

\bibitem{Berenstein:2008jn} 
  D.~Berenstein, R.~Cotta and R.~Leonardi,
  ``Numerical tests of AdS/CFT at strong coupling,''
  Phys.\ Rev.\ D {\bf 78}, 025008 (2008)
  [arXiv:0801.2739 [hep-th]].

\bibitem{Horowitz:1991cd} 
  G.~T.~Horowitz and A.~Strominger,
  ``Black strings and P-branes,''
  Nucl.\ Phys.\ B {\bf 360}, 197 (1991).

\bibitem{Catterall:2007fp} 
  S.~Catterall and T.~Wiseman,
  ``Towards lattice simulation of the gauge theory duals to black holes and hot strings,''
  JHEP {\bf 0712}, 104 (2007)
  [arXiv:0706.3518 [hep-lat]].

\bibitem{Mathur:2005zp} 
  S.~D.~Mathur,
  ``The Fuzzball proposal for black holes: An Elementary review,''
  Fortsch.\ Phys.\  {\bf 53}, 793 (2005)
  [hep-th/0502050].

\bibitem{Brezin:1977sv} 
  E.~Brezin, C.~Itzykson, G.~Parisi and J.~B.~Zuber,
  ``Planar Diagrams,''
  Commun.\ Math.\ Phys.\  {\bf 59}, 35 (1978).

\bibitem{Kokkotas:1999bd} 
  K.~D.~Kokkotas and B.~G.~Schmidt,
  ``Quasinormal modes of stars and black holes,''
  Living Rev.\ Rel.\  {\bf 2}, 2 (1999)
  [gr-qc/9909058].

\bibitem{Berti:2009kk} 
  E.~Berti, V.~Cardoso and A.~O.~Starinets,
  ``Quasinormal modes of black holes and black branes,''
  Class.\ Quant.\ Grav.\  {\bf 26}, 163001 (2009)
  [arXiv:0905.2975 [gr-qc]].
  
\bibitem{Konoplya:2011qq} 
  R.~A.~Konoplya and A.~Zhidenko,
  ``Quasinormal modes of black holes: From astrophysics to string theory,''
  Rev.\ Mod.\ Phys.\  {\bf 83}, 793 (2011)
  [arXiv:1102.4014 [gr-qc]].
  
\bibitem{Kovtun:2005ev} 
  P.~K.~Kovtun and A.~O.~Starinets,
  ``Quasinormal modes and holography,''
  Phys.\ Rev.\ D {\bf 72}, 086009 (2005)
  [hep-th/0506184].

\bibitem{Marolf:2012dr} 
  D.~Marolf and M.~Rangamani,
  ``Causality and the AdS Dirichlet problem,''
  JHEP {\bf 1204}, 035 (2012)
  [arXiv:1201.1233 [hep-th]].
 
\bibitem{Iqbal:2008by} 
  N.~Iqbal and H.~Liu,
  ``Universality of the hydrodynamic limit in AdS/CFT and the membrane paradigm,''
  Phys.\ Rev.\ D {\bf 79}, 025023 (2009)
  [arXiv:0809.3808 [hep-th]].
 
\bibitem{Florido}
R. Florido, J.M. Martin-Gonz\'{a}lez, and J.M. Gomez Llorente,
``Locating Pollicott-Ruelle resonances in chaotic dynamical systems: A class of numerical schemes,"
Phys. Rev. E {\bf 66}, 046208 (2002)
 [arXiv:nlin/0205043 [nlin.CD]].
 
\bibitem{Gaspard-book}
P. Gaspard, {\it Chaos, Scattering and Statistical Mechanics.} Cambridge: Cambridge Univ. Press (1998).



\bibitem{Berenstein:2012ts} 
  D.~Berenstein and E.~Dzienkowski,
  ``Matrix embeddings on flat $R^3$ and the geometry of membranes,''
  Phys.\ Rev.\ D {\bf 86}, 086001 (2012)
  [arXiv:1204.2788 [hep-th]].


\bibitem[Gaspard(2006)]{2006PhyA..369..201G} P. Gaspard, 
``Hamiltonian dynamics, nanosystems, and nonequilibrium statistical mechanics," 
Physica A, 369, 201,
(2006)
[arXiv:cond-mat/0603382 [cond-mat.stat-mech]]


\bibitem{MO87019} F.~Bornemann (mathoverflow.net/users/13034), ``Traceless GUE : Four Centered Fermions," \url{http://mathoverflow.net/questions/87019} (version: 2012-01-30)















\end{thebibliography}
\end{document}